\input mates.sty
\input textos.sty
\input formatos.sty
\font\twelverm=cmr12
\font\sheriff=cmss10 at 9.5pt 

\input epsf
\titpapb{Adel, Barreiro and Yndur\'ain}{Small $x$ deep inelastic scattering}{1}

\rightline{Revised and expanded 30 Dec 1996}
\rightline{FTUAM 96-39\kern0.8em }
\rightline{hep-ph/96 10380}

\vskip1.cm
\centerline{{\twelverm Theory of Small $x$  Deep Inelastic Scattering
 :}}
\vskip.1cm
\centerline{\twelverm NLO Evaluations, and low $Q^2$ Analysis\footnote*{\petit This paper 
includes the results from FTUAM 96-39 [hep-ph 96 10380] and FTUAM 96-44 [hep-ph  96 12469]}}
\vskip1.cm
\centerline{\bf K. Adel,}
\vskip.1cm
\centerline{\bf F. Barreiro}
\vskip.1cm
\centerline{and}
\vskip.1cm
\centerline{{\bf  F. J. Yndur\'ain}\footnote{**}{\petit e-mail: fjy@delta.ft.uam.es}}
\vskip.3cm
\centerline{\sl Departamento de F\'\i sica Te\'orica, C-XI}
\centerline{\sl Universidad Aut\'onoma de Madrid}
\centerline{Canto Blanco, 28049-Madrid}
\vskip1.cm
\setbox0\vbox{\hsize120mm
\noindent{\bf ABSTRACT}.
\vskip.5cm
{\petit We calculate structure functions  at small $x$ both under the assumption of 
a hard singularity (essentially, a power behaviour
 $x^{-\lambda},\,\lambda$ positive, for $x\rightarrow 0$) 
or that of a soft-Pomeron 
dominated behaviour, also called double scaling limit, for the singlet component. A full 
next to leading order (NLO) analysis is carried for the functions $F_2,\,F_{\rm Glue}$ 
and the longitudinal one $F_L$ in $ep$ scattering, and for $x F_3$ 
in neutrino scattering. The results of the calculations are compared
 with experimental data, particularly the recent ones from HERA, in the range 
$x\leq 0.032,\,10\;\gev^2\leq Q^2\leq 1\,500\;\gev^2$. We get reasonable fits,
 with a chi-squared/d.o.f. around two unities, with 
only three-four parameters  for both assumptions, but none of the 
assumptions is by itself able to give a fully satisfactory description of the data.
 The results improve substantially if {\it combining} a soft and a hard component; in this case  
 it is even possible to extend the analysis, phenomenologically, to small values of 
$Q^2$, $0.31\,\gev^2\leq Q^2\leq 8.5\,\gev^2$, and in the $x$ range
 $6\times10^{-6}\lsim x \lsim 0.04$,  
 with the same hard plus soft Pomeron 
hypothesis by assuming  
 a saturating expression for the strong coupling,
 $$\tilde{\alpha}_s(Q^2)=4\pi/\beta_0\log[(Q^2+\Lambdav_{\rm eff}^2)/\Lambdav_{\rm eff}^2].$$ 
The description for low $Q^2$ implies self-consistent values for the parameters 
in the exponents of $x$ both for singlet 
and nonsinglet components. One has to have, for the Regge intercepts, $\alpha_{\rho}(0)=0.48$
 and $\alpha_P(0)=1.470$ [$\lambda=0.470$], 
in uncanny agreement with other determinations of these parameters, and in particular
 the results of the large $Q^2$ fits. The fit to data is so good that we may look (at 
large $Q^2$) for 
signals of a ``triple Pomeron" vertex, for which some evidence is found.

The quality of the calculations of $F_2$, and of the predictions for $F_{\rm Glue},\,F_L$ is only 
marred by the {\it very} large size of the NLO corrections for the singlet part of $F_2$. 
This, in particular, forbids a truly reliable determination of the QCD parameter, $\Lambdav$.}}  
\centerline{\box0}
\vfill\eject
{\sheriff
\centerline{1. {\bf INTRODUCTION}}
\vskip.5cm
In two recent papers$^{[1,2]}$ it was shown how the recent HERA data$^{[3,4]}$
 on electroproduction at small $x$ could be very well fitted by the formulas, proposed long ago 
by C. L\'opez and one of us$^{[5,6]}$ which to leading order (LO) read,
$$F_2(x,Q^2)=\langle e^2_q\rangle[F_S(x,Q^2)+F_{NS}(x,Q^2)],\eqno(1.1{\rm a})$$
$$F_S(x,Q^2)\simeqsub_{x\rightarrow 0}
B_S\alpha_s^{-d_+}x^{-\lambda},\eqno (1.1{\rm b})$$
$$F_{NS}(x,Q^2)\simeqsub_{x\rightarrow 0}
B_{NS}\alpha_s^{-d_{NS}}x^{\rho},\,\rho\sim 0.5.\eqno (1.2)$$
Here $d_+$ and $d_{NS}$ are known quantities related to the singlet 
 anomalous dimension matrix $\bf D$ and to the nonsinglet anomalous dimension; 
$B_S,\,B_{NS},\,\lambda$ are free parameters.
In ref. 2, the analysis was extended to the gluon structure function, while the longitudinal 
one (to LO) had been considered in ref. 8. One has,
$$F_G(x,Q^2)\simeqsub_{x\rightarrow 0}
B_G\alpha_s^{-d_+}x^{-\lambda},\eqno (1.3{\rm a})$$
and
$$R\equiv F_L/F_2\simeqsub_{x\rightarrow 0}r_0\dfrac{\alpha_s}{\pi}=
\frac{C_F\alpha_s}{\pi(2+\lambda)}\bigg\{1+
\frac{4T_Fn_f}{(3+\lambda)C_F}\frac{B_G}{B_S}\bigg\},\eqno (1.3{\rm b})$$
with $B_G/B_S=(d_+-D_{11})/D_{12}$.  
These formulas follow at leading order from perturbative QCD, plus the 
assumption that the leading singularity in $n$ of the matrix elements (for 
 e.g. the singlet case)
$$\langle p|{\bf a}_n| p\rangle;\;
{\bf a}_n=\left(\matrix{\bar{q}_f\gamma\oversetbrace{n-1}\to {\partial\dots\partial} q_f
\cr G\undersetbrace{n-2}\to{\partial\dots\partial} G\cr}\right)$$
are located to the right of those of the corresponding 
anomalous dimensions, $\bf D$, $d_{NS}$ (For more details, see refs. 1, 5, 6, 7).

 An alternate possibility, that may be 
called ``soft-Pomeron" dominated, occurs when the singularity of $\bf D$ 
is the leading one; it was proposed first by De R\'ujula et al.$^{[9]}$ The ensuing LO  
behaviour for the structure functions $F_S,\,F_G$ 
was evaluated in detail by F. Martin$^{[10]}$  and, 
for $R$, in ref. 8. The next to leading order (NLO) corrections to $F_S$ were 
given in ref. 11. To LO one has, 
$$\eqalign{F_S\simeq
 \frac{c_0}{|\log x|}\left[ \frac{9|\log x| \log[\alpha_s(Q_0^2)/\alpha_s (Q^2)]}{4\pi^2(33-2n_f)}
\right]^{\frac{1}{4}}\cr
\times\exp\left\{ \sqrt{d_0|\log x|\;
\left[\log\frac{\alpha_s(Q_0^2)}{\alpha_s(Q^2)}\right]}-
d_1\log\frac{\alpha_s(Q_0^2)}{\alpha_s(Q^2)} \right\},}
\eqno (1.4{\rm a})$$
$$\eqalign{F_G\simeq
 \dfrac{9c_0}{n_f}\left[ \frac{33-2n_f}{576\pi^2|\log x| \log[\alpha_s(Q_0^2)/\alpha_s (Q^2)]}
\right]^{\frac{1}{4}}\cr
\times\exp\left\{\sqrt{d_0|\log x|\;
\left[\log\frac{\alpha_s(Q_0^2)}{\alpha_s(Q^2)}\right]}-
d_1\log\frac{\alpha_s(Q_0^2)}{\alpha_s(Q^2)}\right\},}
 \eqno(1.4{\rm b})$$
$d_0=144/(33-2n_f),\,d_1=(33+2n_f/9)/(33-2n_f)$, and
$$R\simeq
 \left[\frac{(33-2n_f)|\log x|}{\log[\alpha_s(Q_0^2)/\alpha_s (Q^2)]}\right]^{\frac{1}{2}}
\;\frac{\alpha_s (Q^2)}{2\pi}, \eqno(1.4{\rm c})$$
to be compared with Eqs. (1.1), (1.3). We will discuss this possibility 
in Sect. 4.

Returning to our first case, the analysis was extended in ref. 2 to practically
 the whole range of HERA data at the cost of introducing phenomenological correction 
terms for ``large" values of $x$, $x>0.01$, and small $Q^2<12\;\gev^2$. Moreover, only the 
LO prediction for $R$ was evaluated. Here we go a few steps forward, in the following 
directions. First, we perform a full NLO analysis, including that of the longitudinal 
structure function. Second, we extend the analysis to incorporate 
the function $xF_3$ for neutrino 
scattering which provides pure nonsinglet function, hence a  combination independent
 from that in (1.1a): this helps stabilize the results.  
Then, for the hard Pomeron case, 
 we include theoretically justified corrections, and resummations,
 which enable us to extend the 
range of validity of the formulas (as determined e.g. in ref. 1) and to 
paliate somewhat the effects of the hughe size of the 
NLO corrections to the singlet component of $F_2$, which 
are of some $\sim 6\alpha_s$.

For the soft Pomeron dominance hypothesis we again perform a full NLO calculation 
 of $F_S,\,F_G$ and $F_L$. NLO corrections are, also in 
this case, very large; 
 their size is indeed the only serious drawback for 
 our results, otherwise able to describe reasonably  
well the HERA data in a wide range, and providing believable predictions for 
the gluon and longitudinal structure functions. This agreement of 
theoretical predictions and data is essentially true both for the soft and hard Pomeron 
hypotheses, an apparently surprising fact that is discussed in Sect. 5.
 Here we argue that a possible reason 
is that $F_S$ contains {\it both} a hard and a soft piece. In fact, we are 
able to give excellent fits to all HERA data ($x\leq 0.032$) by using a formula sum of (1.1) 
and (1.4). This provides us with the best fits to 
the data, the parameters of which are reported in Table X.

 In \sect 6 we show that the fit 
with a soft plus a hard Pomeron may 
be phenomenologically extended to low $Q^2$, down to 0.31 GeV$^2$, provided we 
make a saturation assumption for the strong coupling $\alpha_s$ and satisfy a 
self-consistency condition for the 
parameters $\lambda,\,\rho$. This fixes
 these parameters to the values $\lambda=0.470,\,\rho=0.522$, in uncanny agreement 
with other (in particular high $Q^2$, \sect 5) determinations.    
\vskip.4cm
\centerline{2. {\bf THEORETICAL EVALUATIONS (HARD SINGULARITY)}}
\vskip0.5cm
We will here
 briefly rederive the extension to NLO of the equations
 governing the behaviour of structure functions 
as $x\rightarrow 0$; not only for 
ease of reference, but because the large size of the {\it singlet} NLO 
corrections makes it convenient to use formulas more precise than those employed 
in refs. 2, 6. We will also extend the analysis to the longitudinal 
structure function at NLO.
\vskip.1cm
{\bf 2.1. Nonsinglet}
\vskip.2cm
Defining the moments
$$\mu_{NS}(n,Q^2)=\int^1_0\dd x\,x^{n-2}F_{NS}(x,Q^2),\eqno (2.1)$$
they satisfy the QCD evolution equations
$$\mu_{NS}(n,Q^2)=
\ee^{-\int^t_{t_0}\dd t'\,\gamma_{NS}(n,g(t'))}C_{NS}(n,\alpha_s(Q^2)),\eqno (2.2)$$
where $t=\log Q^2$, $g$ is the coupling constant, $\gamma_{NS}$ the nonsinglet (NS)  
anomalous dimension, and $C_{NS}$ the NS Wilson 
coefficient\fonote{The anomalous dimensions and coefficients are collected in the 
Appendix for ease of reference.}. The first 
singularity of $\gamma_{NS}$ to LO and NLO lies at $n=0$. If we assume 
this to occur to all orders, and that $F_{NS}$ behaves like a power $x^{\rho}$ 
as $x$ goes to zero, we get the following behaviour from (2.2):
$$F_{NS}(x,Q^2)\simeqsub_{x\rightarrow 0}
({\rm Const.})\,x^{\rho}\,\ee^{-\int^t_{t_0}\dd t'\,\gamma_{NS}(1-\rho,g(t'))}
C_{NS}(1-\rho,\alpha_s(Q^2)).
\eqno (2.2)$$
Expanding $\gamma_{NS},\,g,\,C_{NS}$ to second order and integrating 
we get,
$$F_{NS}(x,Q^2)\simeqsub_{x\rightarrow 0}
B_{NS}x^{\rho}\left[1+\frac{C_{NS}^{(1)}(1-\rho)\alpha_s(Q^2)}{4\pi}+\dots\right]$$
$$\times \exp\left\{d_{NS}(1-\rho)\log \alpha_s^{-1}(Q^2)+
\frac{q_{NS}(1-\rho)\alpha_s(Q^2)}{4\pi}+\dots\right\},\eqno (2.4)$$
$B_{NS}$ a constant and
$$d_{NS}(n)=-\gamma^{(0)}_{NS}(n)/2\beta_0$$
$$q_{NS}(n)=\frac{\beta_1 d_{NS}(n)}{\beta_0}+\frac{\gamma^{(1)}_{NS}(n)}{2\beta_0}.$$
The values of the quantities $\beta_i,\,C_{NS},\,\gamma_{NS}^{(i)}$ may be found in refs. 6, 7. 
Eq. (2.4) may be conveniently rewritten (suppressing the dots) as
$$\eqalign{F_{NS}(x,Q^2)\simeqsub_{x\rightarrow 0}
B_{NS}\left\{1+\frac{C_{NS}^{(1)}(1-\rho)\alpha_s(Q^2)}{4\pi}\right\}\cr
\times\ee^{q_{NS}(1-\rho)\alpha_s/4\pi}\left[\alpha_s(Q^2)\right]^{-d_{NS}(1-\rho)}x^{\rho},}
\eqno (2.5)$$
an expression\fonote{Eq. (2.5) corrects the sign misprints in Eq. (2.18) of ref. 6.} 
in which the modifications of (1.2) due 
to the NLO corrections is apparent.

 One can also expand the exponent in (2.5) and get
$$F_{NS}(x,Q^2)\simeqsub_{x\rightarrow 0}
B_{NS}\left\{1+\frac{v_{NS}(1-\rho)\alpha_s(Q^2)}{4\pi}\right\}\alpha_s^{-d_{NS}}x^{\rho},
\eqno (2.6)$$
$v_{NS}=C_{NS}^{(1)}+q_{NS}$.

For $\rho=0.5$, a value that follows from 
a Regge analysis and that we will adopt here, one finds
$$v_{NS}|_{\rho=0.5,n_f=4}=3.42,\eqno (2.7)$$
so the NLO correction is small and we may use (2.5) or (2.6) indifferently.   
\vskip.1cm
{\bf 2.2. Singlet}
\vskip.2cm
Eq. (2.2) is now replaced by the coupled equations
$$\ybf{\mu}(n,Q^2)=({\rm Const.}){\rm T}\;
\ee^{-\int^t_{t_0}\dd t'\,\ybf{\scriptstyle\gamma}(n,g(t'))}{\bf C}(n,\alpha_s(Q^2))\eqno (2.8)$$
$$\ybf{\mu}(n,Q^2)=\left(\matrix{\mu_S(n,Q^2)=\int^1_0\dd x\,x^{n-2}F_S(x,Q^2)\cr
\mu_G(n,Q^2)=\int^1_0\dd x\,x^{n-2}F_G(x,Q^2)\cr}\right).$$
Here $\ybf{\gamma},\,{\bf C}$ are square matrices;
 the operation $T$ in (2.8) is like the familiar 
time ordering operator but it now orders in $t=\log Q^2$.
This ordering, and the matrix character of the equations 
complicates the singlet analysis, the details of which
 may be found in refs. 6, 7. To {\it next} to 
leading order we may easily write the analogue of (2.4), (2.5) as
$$F_S(x,Q^2)\simeqsub_{x\rightarrow 0}
B_Sx^{-\lambda}\left[1+\frac{c_S(1+\lambda)\alpha_s(Q^2)}{4\pi}+\dots\right]$$
$$\times \exp\left\{d_+(1+\lambda)\log \alpha_s^{-1}(Q^2)+
\frac{q_S(1+\lambda)\alpha_s(Q^2)}{4\pi}+\dots\right\},\eqno (2.9)$$
or, suppressing the dots and in a form easier to compare with 
the LO expression (1.1b),
$$\eqalign{F_S(x,Q^2)\simeqsub_{x\rightarrow 0}
B_S\left\{1+\frac{c_S(1+\lambda)\alpha_s(Q^2)}{4\pi}\right\}\cr
\times\ee^{q_S(1+\lambda)\alpha_s/4\pi}\left[\alpha_s(Q^2)\right]^{-d_+(1+\lambda)}x^{-\lambda}.}
\eqno (2.10{\rm a})$$
A corresponding equation for the gluon
 component we will consider later. In above equations we have
$${\bf D}=-\frac{\ybf{\gamma}^{(0)}}{2\beta_0},$$
$$c_S=C^{(1)}_{11}+\frac{d_+-D_{11}}{D_{12}}\,C^{(1)}_{12},$$
$$q_S=\frac{\beta_1d_+}{\beta_0}+\frac{\bar{\gamma}_{11}}{2\beta_0}+
\frac{\bar{\gamma}_{21}}{2\beta_0(d_- -d_++1)}\;\frac{D_{12}}{d_--d_+},$$
$$\bar{\ybf{\gamma}}={\bf S}^{-1}\ybf{\gamma}^{(1)}{\bf S},$$
and ${\bf S}$ is the matrix that diagonalizes ${\bf D}$:
$${\bf S}^{-1}{\bf D}{\bf S}=\left(\matrix{d_+&0\cr 0&d_-\cr}\right).$$
One can also expand the exponential in (2.10{\rm a}) 
and get
$$F_S(x,Q^2)\simeqsub_{x\rightarrow 0}
B_S\left\{1+\frac{w_S(1+\lambda)\alpha_s(Q^2)}{4\pi}\right\}\alpha_s^{-d_+}x^{-\lambda},
\eqno (2.10'{\rm a})$$
$w_S=c_S+q_S$.
The $\bar{\gamma},\,q,\,c,\,w$ are collected in the Appendix.

Unfortunately, $w_S$ is very large. For $\lambda=0.35$, $n_f=4$, 
$w_S=77.8$; for $\lambda=0.47$, $w_S=56.7$. Therefore, we are faced with the 
 choice of using the exponential form 
(2.10) or the expanded one (2.$10'$). The 
exponential form has errors of order $\alpha_s^2$ because the noncommutativity 
of $\ybf{\gamma}^{(0)},\,\ybf{\gamma}^{(1)}$ makes the $T$-exponential different 
from the ordinary exponential. If we use the expanded form (2.$10'$) we have 
other errors (also  of order $\alpha_s^2$) due to the large size of 
the neglected term ${\rm O}[q_S(1+\lambda)\alpha_s/4\pi]^2$. It is unclear a 
priori which of the 
two procedures will be more 
accurate, although the abnormally large size of $q_S(1+\lambda)$
 suggests that the exponentiated form will be more precise; note that the
 perturbative expansion still makes sense, {\it for the exponent}, in 
that for reasonably large $Q^2$ one has
$$\frac{q_S\alpha_s}{4\pi}\ll d_+\,\log \alpha_s^{-1}.$$
In fact, and as we will see, the exponentiated form produces somewhat more satisfactory 
results than the expanded one. 
At any rate, we will use both (2.10) and (2.$10'$): one may take
 the difference as an indication of 
 the {\it theoretical} error of our calculation.

 Similar considerations of course apply to the 
gluon component that we discuss next, although in this case the correction is 
much smaller ($\sim 15\alpha_s/4\pi$) so use of exponentiated 
or expanded form is essentially equivalent here and only
 consistency with the quark component will 
make us use one or the other. We then have,

$$\eqalign{F_G(x,Q^2)\simeqsub_{x\rightarrow 0}
 B_S\frac{d_+-D_{11}}{D_{12}}\left\{1+\frac{c_G(1+\lambda)\alpha_s(Q^2)}{4\pi}\right\}\cr
\times\ee^{q_G(1+\lambda)\alpha_s/4\pi}\left[\alpha_s(Q^2)\right]^{-d_+(1+\lambda)}x^{-\lambda},}
\eqno (2.10{\rm b})$$
where now
$$c_G=C^{(1)}_{11}+
C^{(1)}_{12}\left(\frac{D_{22}-D_{11}}{D_{12}}+\frac{D_{21}}{d_+-D_{11}}\right)\equiv c_S,$$
$$q_G=\frac{\beta_1d_+}{\beta_0}+\frac{\bar{\gamma}_{11}}{2\beta_0}+
\frac{\bar{\gamma}_{21}}{2\beta_0(d_- -d_++1)}\;\frac{D_{12}}{d_--d_+}\;
\frac{d_--D_{11}}{d_+-D_{11}}.$$
In expanded form,
$$\eqalign{F_G(x,Q^2)\simeqsub_{x\rightarrow 0}
B_S\frac{d_+-D_{11}}{D_{12}}
\left\{1+\frac{w_G(1+\lambda)\alpha_s(Q^2)}{4\pi}\right\}
\left[\alpha_s(Q^2)\right]^{-d_+(1+\lambda)}x^{-\lambda},\cr
 w_G(1+\lambda)\equiv c_G(1+\lambda)+q_G(1+\lambda).\kern10em\phantom{x}}
\eqno (2.10'{\rm b})$$
\vskip.1cm
{\bf 2.3. The longitudinal structure function}
\vskip.2cm
We normalize the longitudinal structure function $F_L$ in 
such a way that one has 
$$R(x,Q^2)=\frac{F_L}{F_S+F_{NS}-F_L}. \eqno (2.11{\rm a})$$
It is also convenient to define the quantity $R'$ by
$$R'(x,Q^2)=\frac{F_L}{F_S+F_{NS}};\;R=\frac{R'}{1-R'}. \eqno (2.11{\rm b})$$
 For $x\rightarrow 0$ the contribution of 
$F_{NS}$ is negligible with respect to that of $F_S$ and we will 
accordingly neglect it; the effect of taking it into account, to LO, may be 
found in ref. 8. The function $F_L$ may be 
evaluated in terms of $F_S,\,F_G$. One has, 
$$F_L(x,Q^2)=\int^1_x\dd y\,\left\{C^L_{S}(y,Q^2)F_S\left(\frac{x}{y},Q^2\right)+
C^L_{G}(y,Q^2)F_G\left(\frac{x}{y},Q^2\right)\right\},\eqno (2.12{\rm a})$$
where the kernels $C^L$ are,
$$\eqalign{C^L_{S}(x,Q^2)=C^L_{NS}(x,Q^2)+C^L_{PS}(x,Q^2);\cr
C^L_{NS}(x,Q^2)=[4C_Fx]\frac{\alpha_s(Q^2)}{4\pi}+
c^{(1)L}_{NS}(x)\left(\frac{\alpha_s(Q^2)}{4\pi}\right)^2+\dots\cr
C^L_{PS}(x,Q^2)=c^{(1)L}_{PS}(x)\left(\frac{\alpha_s(Q^2)}{4\pi}\right)^2+\dots \cr
C^L_{G}(x,Q^2)=[16n_fT_Fx(1-x)]\frac{\alpha_s(Q^2)}{4\pi}+
c^{(1)L}_G(x)\left(\frac{\alpha_s(Q^2)}{4\pi}\right)^2+\dots\,.}\eqno (2.12{\rm b})$$
The functions $c^{(1)L}_{NS,PS,G}$ are described in 
the Appendix\fonote{These quantities were first evaluated in refs. 12, 13. Correct values, 
checked at least in two independent calculations, are given in Eqs. (8), (9) of 
ref. 12 for $c^{(1)L}_{NS,PS}$ and in Eq. (9) of ref. 13 for $c^{(1)}_G$. The 
value of this quantity given in ref. 12 contains an error. The 
first moments of the $c$ may be found in ref. 14; they are 
useful, among other things, to check the integrals (2.15) here.}.

Under our assumptions, Eqs. (2.10), the behaviour 
of $F_L$ follows immediately; we have,
$$F_L(x,Q^2)\simeqsub_{x\rightarrow 0}
F_S(x,Q^2)\,\int^1_0\dd x\,x^{\lambda}\left\{C^L_{S}(x,Q^2)+\eta(Q^2)C^L_{G}(x,Q^2)\right\},
\eqno (2.13)$$
and $\eta$ is the ratio 
$$\eta(Q^2)=\lim_{x\rightarrow 0}\frac{F_G(x,Q^2)}{F_S(x,Q^2)}.$$

To LO a simple evaluation gives$^{[8]}$ $R(x,Q^2)\simeq R^{(0)}(x,Q^2)$ with
$$\eqalign{ R^{(0)}(x,Q^2)\simeqsub_{x\rightarrow 0}r_0(1+\lambda)\frac{\alpha_s}{\pi},\cr
r_0(1+\lambda)=\frac{C_F\alpha_s}{2+\lambda}\left\{1+\frac{4T_Fn_f}{(3+\lambda)C_F}\;
\frac{d_+(1+\lambda)-D_{11}(1+\lambda)}{D_{12}(1+\lambda)}\right\}.}\eqno (2.14)$$
To NLO the calculation is made numerically. For this, define the integrals (whose 
 values may be found in the Appendix),
$$\eqalign{\int_0^1\dd x\,x^{\lambda}\,c^{(1)L}_{NS}(x)=
C_F(C_A-2C_F)I_{S1}(\lambda)+C_F^2I_{S2}(\lambda)+C_FT_Fn_fI_{S3}(\lambda),\cr
\int_0^1\dd x\,x^{\lambda}\,c^{(1)L}_{PS}(x)=C_FT_Fn_fI_{PS}(\lambda),\cr
\int_0^1\dd x\,x^{\lambda}\,c^{(1)L}_G(x)=n_fT_F[C_FI_{G1}(\lambda)+C_AI_{G2}(\lambda)],}
\eqno (2.15)$$
and write, in exponentiated form,
$$\eta(Q^2)=\frac{d_+-D_{11}}{D_{12}}\ee^{(q_G-q_S)\alpha_s/4\pi}.\eqno (2.16{\rm a})$$
Then,
$$R'=R^{(0)}+R'^{(1)}\eqno (2.16{\rm b})$$
with $R^{(0)}$ as above and
$$\eqalign{R'^{(1)}\simeqsub_{x\rightarrow 0}\frac{C_F}{2+\lambda}\;\frac{\alpha_s(Q^2)}{\pi}
\left\{1+
\frac{2+\lambda}{4}
\left[C_AI_{S1}+C_F(I_{S2}-2I_{S1})+n_fT_F(I_{S3}+I_{PS})\right]\frac{\alpha_s}{4\pi}\right.\cr
\left.+\frac{4n_fT_F}{(3+\lambda)C_F}\left[1+
\frac{(2+\lambda)(3+\lambda)}{16}(C_FI_{G1}+C_AI_{G2})\frac{\alpha_s}{4\pi}\right]\eta(Q^2)\right\}.
}\eqno (2.16{\rm c})$$ 
\vskip.4cm
\centerline{ 3. {\bf NUMERICAL RESULTS (HARD SINGULARITY ONLY).}}
\vskip.5cm
{\bf 3.1. The function $F_2$.}
\vskip.2cm
{\sl LO calculations}.
 For ease of comparison between LO and NLO evaluations we repeat
 here the results of a fit to the old (1993) 
Zeus data$^{[3]}$, as performed in ref. 1. The calculation is carried for 32 points in the range 
$$x<10^{-2},\;12\leq Q^2\leq 90\;\gev^2.$$
Because of the size of the experimental errors a LO calculation 
is sufficient, and the NS contribution may be neglected. The QCD parameter
 $\Lambdav$ is fixed to $0.2\;\gev$ so that $\alpha_s(m_{\tau}^2)=0.32$. The results are 
summarized in Table I for $n_f=4$ 
flavours. The corresponding values of $B_G,\; r_0$ are also given.
 The agreement of the value of $\lambda$ with the figure 
 $\lambda=0.36\pm0.07$ obtained in ref. 3 
from data with $x\geq 0.02,\,Q^2\leq 22\,\gev^2$ is noteworthy.
\vskip.2cm
\setbox0=\vbox{\hsize=14cm
{\petit
\vskip.2cm
\centerline{{\bf Table I}. ``Old" Zeus data. LO.  $n_f=4;\Lambdav(1\,{\rm loop},n_f=4)=0.200\,{\rm GeV};
\,\alpha_s(m_{\tau}^2)=0.32.$}
\vskip.1cm
\centerline{\kern16em\hrulefill\kern16em}
$$\matrix
{\lambda & d_+ & \langle e_q^2\rangle B_S& B_G/B_S&r_0&\chi^2/{\rm d.o.f.}\cr
 & & &\cr
0.38\pm0.01&2.41\pm0.1&(2.70\pm0.22)\times 10^{-3}&20.56\pm0.54&6.24\pm 0.24&\tfrac{9.13}{32-2}\cr
}$$
}
\vskip.1cm}
\centerline{\boxit{\box0}}

{\sl  NLO evaluation}. If we only fit the H1 points\ref{4} with $x<10^{-2}$, $Q^2\geq 12\;\gev^2$
 using the 
exponentiated formulas, we get a $\chi^2$/d.o.f.
 is less than one, with parameters 
reported in Table II.
\vskip.2cm
\setbox0=\vbox{\hsize=12cm
{\petit
\vskip.2cm
\centerline{{\bf Table II}. H1 plus $\nu$ data. $x\leq .01$, $n_f=4$; $\Lambdav(n_f=4,\,$ 2 loop) fixed to 0.11$\; \gev$.}
\vskip.1cm
\centerline{\kern16em\hrulefill\kern16em}
$$\matrix
{\lambda & \langle e_q^2\rangle B_S& \langle e_q^2\rangle B_{NS}&\chi^2/{\rm d.o.f.}\cr
& & &\cr
0.3218&1.423\times 10^{-4}&0.390&\tfrac{48.9}{58-3}\cr}$$
}
\vskip.1cm}
\centerline{\boxit{\box0}}

 However, 
it is still not possible to give any value for the QCD parameter $\Lambdav$. The reason 
is that the interplay between singlet and nonsinglet parts 
compensates the effect of varying $\Lambdav$. For example, a $\chi^2/{\rm d.o.f.}$ less than 
one is attained for $3\;\mev\leq\Lambdav\leq 260\;\mev$. This is  
why we do not give errors in the parameters in Table II.

\setbox4=\vbox{\hsize 7.9truecm
\vskip.2cm  
We may improve the situation as follows. First, and 
as discussed in ref. 1, we can include more points 
limited by a certain $Q^2(x)$ beyond which corrections to the leading
 behaviour become important. 
To be precise, we choose the H1 points with (Fig. 1)
  $$\matrix{Q^2\leq 150\;\gev^2,&{\rm for}\;x=0.013,\cr
Q^2\leq 90\;\gev^2,&{\rm for}\;x=0.02,\cr
Q^2\leq 60\;\gev^2,&{\rm for}\;x=0.032,}\eqno (3.1)$$
with a total of 77 points. 
Secondly, we incorporate small $x$ data (a total of 10 points) from the {\it neutrino} structure
 function$^{[15]}$ $xF_3$ which 
is pure nonsinglet and hence provides the independent measurement necessary to disentangle the 
singlet and nonsinglet components of $F_2$: this, as we will see, gives stability to 
the results.

 The outcome of the fits is given 
in Tables III, IV with $\Lambdav$
 a free parameter. The $\chi^2$/d.o.f. is reasonable, although its increase beyond 
unity reflects the fact that the 
subleading effects are  substantial for the points $x=0.013\,\sim\,0.032$.}
\setbox1=\vbox{\epsfxsize=7.5truecm \epsfbox{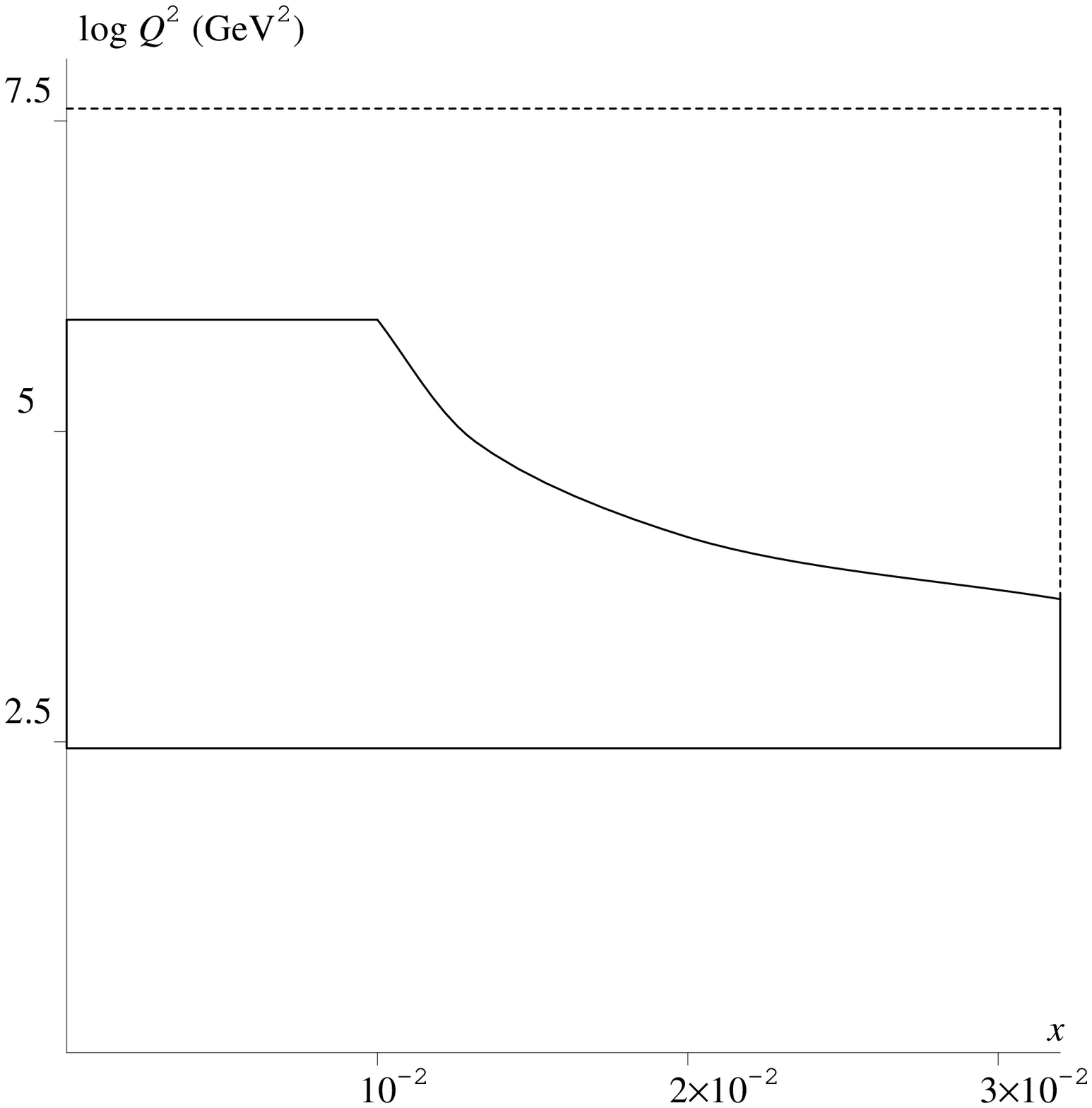}}
\setbox3=\vbox{\hsize62mm\petit \noindent Figure 1. Area described 
by (3.1), bounded by a continuous line. Broken line: 
extended region for fit with (3.2). \hb\vskip.2cm}
\setbox0=\vbox{\hsize=6.8truecm\box1\vskip.2cm\rightline{\box3}}
\line{\box4\hfil\box0}
\vskip.2cm
\setbox0=\vbox{\hsize=14cm
{\petit    
\vskip.2cm
\centerline{{\bf Table III}. H1 plus $\nu$ data; $x$ given by (3.1). $n_f=4$, two loops.}
\vskip.1cm
\centerline{\kern16em\hrulefill\kern16em}
$$\matrix
{\Lambdav & \lambda & \langle e^2_q\rangle B_S& \langle e^2_q\rangle  B_{NS}&\chi^2/{\rm d.o.f.}\cr
 & & & &\cr
0.080_{-0.035}^{+0.060}\,\gev&
0.3243\pm 0.0065&1.321_{-0.427}^{+0.630}\times 10^{-4}&0.254^{+0.025}_{-0.020}
&\tfrac{97.5}{87-4}\cr}$$
\vskip.2cm
\hrule
\vskip.2cm
\centerline{{\bf Table IV}. H1 plus $\nu$ data; $x$ given by (3.1). $n_f=5$, two loops.}
\vskip.1cm
\centerline{\kern16em\hrulefill\kern16em}
$$\matrix
{\Lambdav(n_f=4,\; 2\;{\rm loops}) & \lambda & \langle e^2_q\rangle  B_S& \langle e^2_q\rangle  B_{NS}&\chi^2/{\rm d.o.f.}\cr
 \cr
0.11_{-0.05}^{+0.06}\,\gev&0.331\pm 0.006&1.118_{-0.409}^{+0.444}\times 10^{-4}&0.311^{+0.024}_{-0.072}
&\tfrac{103.0}{87-4}\cr}$$
}
\vskip.1cm}
\centerline{\boxit{\box0}}
From these results it is clear that the data do not discriminate between $n_f=4,\,5$, 
although the first value is slightly favoured. For this reason we will give almost 
exclusively fits with $n_f=4$. 

The values of $\Lambdav$ we obtain are compatible 
with standard ones$^{[16]}$, albeit on the small side. Because the parameters 
are very strongly correlated the errors given are obtained {\it not} by varying 
the parameters independently, but by varying only $\Lambdav$ and treating the other 
 parameters as dependent quantities. It is also important to realize that
 the errors in Tables III, IV and indeed in practically all the 
evaluations, are purely {\it nominal} in the sense that we have not 
taken into account {\it theoretical} errors, which are 
much larger. In fact, the central values for the 
parameters, especially $B_S,\,\lambda$, depend very strongly on the theoretical 
assumptions\fonote{This is in fact the reason 
why we have given a variety of evaluations, and not just the best ones: to get a flavour 
for the systematic theoretical uncertainties.} made; for example, they vary way beyond the 
nominal errors from LO to NLO: compare e.g. Table I with Table II.
 
The results reported in  Tables III, IV were 
obtained with the exponentiated formula. If we use the expanded one, 
Eq. (2.$10'$a) we find the results of Table ${\rm IV}'$.
\vskip.2cm
\setbox0=\vbox{\hsize=13cm
{\petit
\vskip.2cm
\centerline{{\bf Table ${\bf IV}'$}. H1 plus $\nu$ data; $x$ given by (3.1).
 $n_f=4$; $\Lambdav(n_f=4,\; 2\;{\rm loops})=0.080\,\gev$.}
\vskip.1cm
\centerline{\kern16em\hrulefill\kern16em}
$$\matrix
{ \lambda & \langle e^2_q\rangle  B_S& \langle e^2_q\rangle  B_{NS}&\chi^2/{\rm d.o.f.}\cr
 &&&\cr
0.3408&2.54\times 10^{-4}&0.268&\tfrac{99.5}{87-4}}$$
}
\vskip.1cm
}
\centerline{\boxit{\box0}}

We do not give errors, but the results of the 
fit for a representative value of $\Lambdav$: that for which the fit with the  
{\it exponentiated} formula is optimum. This is because there 
is no optimum reasonable value of $\Lambdav$ if using the nonexponentiated 
expression; the $\chi^2$ decreases slowly with $\Lambdav$ down to a few $\mev$.    

{\sl Corrections}. Let us now turn to the corrections 
that will enable us to extend the calculation to {\it all}
 points with $x\leq 0.032,\;Q^2\geq 12\,\gev^2$ (Fig. 1). We take them into account
 semi-phenomenologically\fonote{The similitudes and 
differences with the more phenomenological procedure of ref. 2 should be apparent.}
 by replacing (2.10{\rm a}) with
$$F_S(x,Q^2)\simeq B_S\left\{1+\frac{c_S\alpha_s}{4\pi}\right\}
\ee^{q_S\alpha_s/4\pi}\,\alpha_s^{-d_+}\,x^{-\lambda}(1-x)^{\nu(Q^2)},\eqno (3.2{\rm a})$$
and fixing $\nu(Q^2)$ so that, for small $Q^2$, we agree with the 
result of the counting rules for $x\rightarrow 1$ and, for large $Q^2$, we 
satisfy the momentum sum rule,
$$\int^1_0\dd x\,F_S(x,Q^2)\rightarrowsub_{Q^2\rightarrow \infty}\;\frac{3n_f}{3n_f+16}.$$
Specifically, we choose
$$\eqalign{\nu(Q^2)=\nu_0+
\left\{\frac{B_S[1+c_S\alpha_s/4\pi]
\ee^{q_S\alpha_s/4\pi}
\Gammav(1+\lambda)[16+3n_f]}{3n_f\alpha_s^{d_+}}\right\}^{1/(1-\lambda)},\cr
\nu_0=7.\kern13.5em\phantom{x}}\eqno (3.2{\rm b})$$
Note that this does {\it not} introduce any new parameter.

For $F_{NS}$ we replace (2.6) by
$$F_{NS}\simeq B_{NS}\left\{1+\frac{v_{NS}\alpha_s}{4\pi}\right\}
\alpha_s^{-d_{NS}}x^{\rho}(1-x)^{\nu_{NS}},\eqno (3.2{\rm c})$$
but, because the NS component is only relevant at small values of $Q^2$ we 
fix $\nu_{NS}=3$ independent of $Q^2$ (actually, the $\chi^2$ varies by less 
than one unit for $0\leq \nu_{NS}\leq 4$). Then, we still write 
$F_2= \langle e^2_q\rangle [F_S+F_{NS}]$, 
and, for neutrino scattering, 
$$xF_3=F_{NS}^{\rm odd}(x,Q^2)=B_{NS}\left\{1+\frac{v^{\rm odd}_{NS}\alpha_s}{4\pi}\right\}
\alpha_s^{-d_{NS}}x^{\rho}(1-x)^{\nu_{NS}}.\eqno (3.2{\rm d})$$

\setbox0=\vbox{\hsize 14.7truecm \epsfxsize=14.6truecm\epsfbox{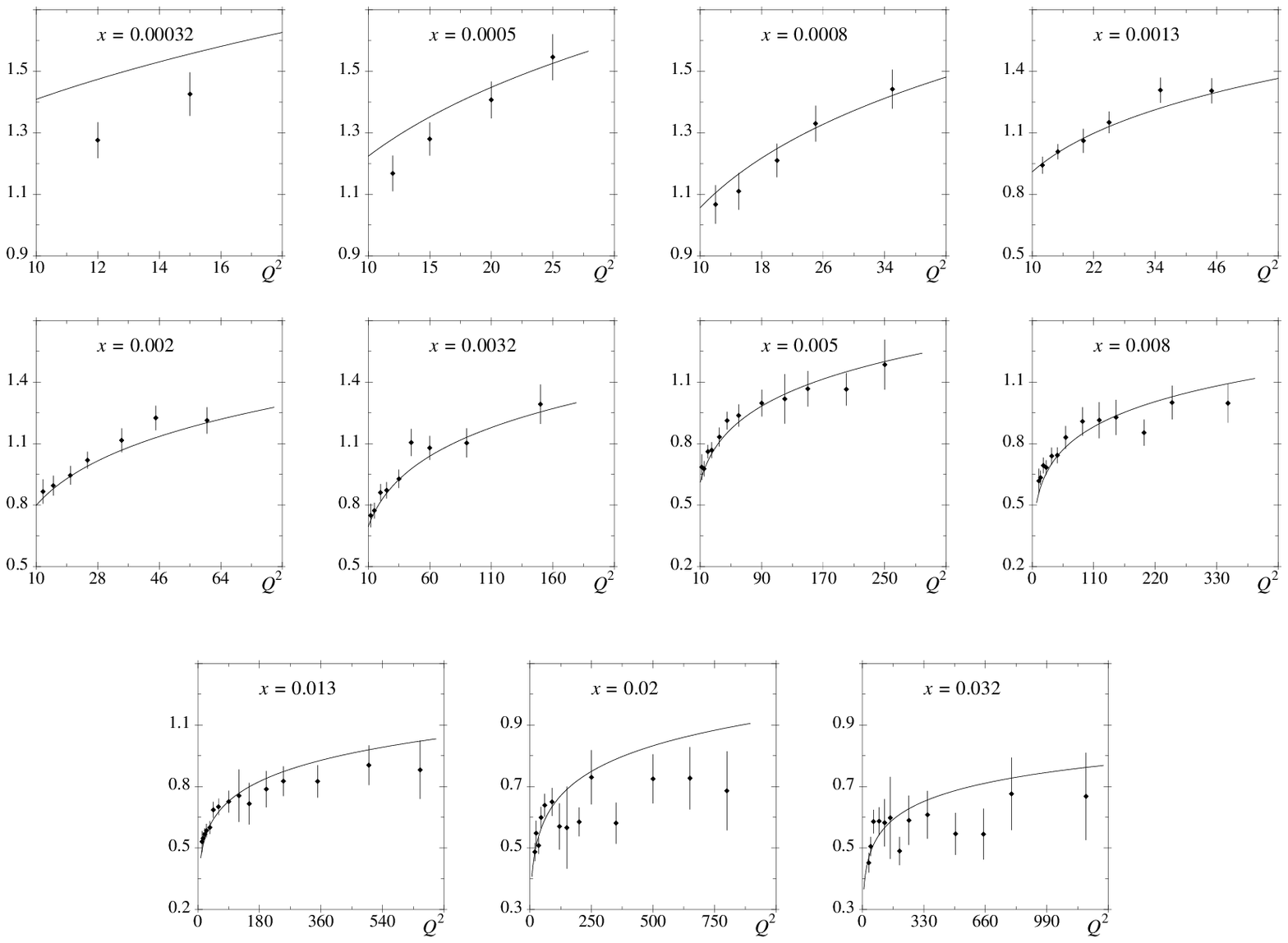}\hb{\petit
\phantom{x}\kern2truecm Figure 2a. Comparison of predictions from 
Eqs. (3.2a, c), Table V
\hb\phantom{x}\kern2truecm ($\Lambdav=0.14\;\gev$), with H1 $ep$ data$^{[4]}$ for $F_2$.
\vskip.2cm}}
\centerline{\box0}
The results of the fit are presented in Table V, for the exponentiated 
expression, Eq. (3.2a). There is unfortunately no minimum 
as a function of $\Lambdav$: the $\chi^2$ decreases slowly with $\Lambdav$. We 
thus give results only for two representative values of this parameter. The 
pictorial representation of the fit is given in Fig. 2a (for $ep$) and Fig. 2b for 
neutrino scattering, both for $\Lambdav=0.14\,\gev$.
\vskip.2cm
\setbox0=\vbox{\hsize=14cm
{\petit
\vskip.2cm
\centerline{{\bf Table V}. H1 plus $\nu$ data. 
 $n_f=4; x\leq 0.032,\;12\;\gev^2\leq Q^2\leq 1\,200\;\gev^2 $.}
\vskip.1cm
\centerrule{16em}
$$\matrix
{ & \lambda & \langle e^2_q\rangle  B_S& \langle e^2_q\rangle  B_{NS}&\chi^2/{\rm d.o.f.}\cr
& & & & \cr
 \Lambdav=
0.10\;\gev:&0.3183&1.292\times 10^{-4}&0.328&\tfrac{127.2}{110-3}\cr
 \Lambdav=
0.20\;\gev:&0.3286&2.257\times 10^{-4}&0.371&\tfrac{141.3}{110-3}}$$
}
\vskip.1cm}
\centerline{\boxit{\box0}}

\setbox1=\vbox{\hsize6.9cm  
The $\chi^2/{\rm d.o.f.}$\ is slightly larger than one. Part 
of the discrepancy is due to the data, some of 
which is clearly 
incompatible with the rest. Also, one may substantially  
improve the $\chi^2$ if introducing a free parameter in the definition of $\nu(Q^2)$, as 
shown e.g. in ref. 2. However, part 
of the disagreement is certainly due to 
rigidity of the theoretical formulas, 
and  to true deviation 
from the model which occur for ``large" values of $x$. We will discuss this further in 
connection with the analysis of the Zeus data, and in Sect. 5.}

\setbox12=\vbox{\hsize6.6truecm \centerline{{\epsfxsize=3.3truecm\epsfbox{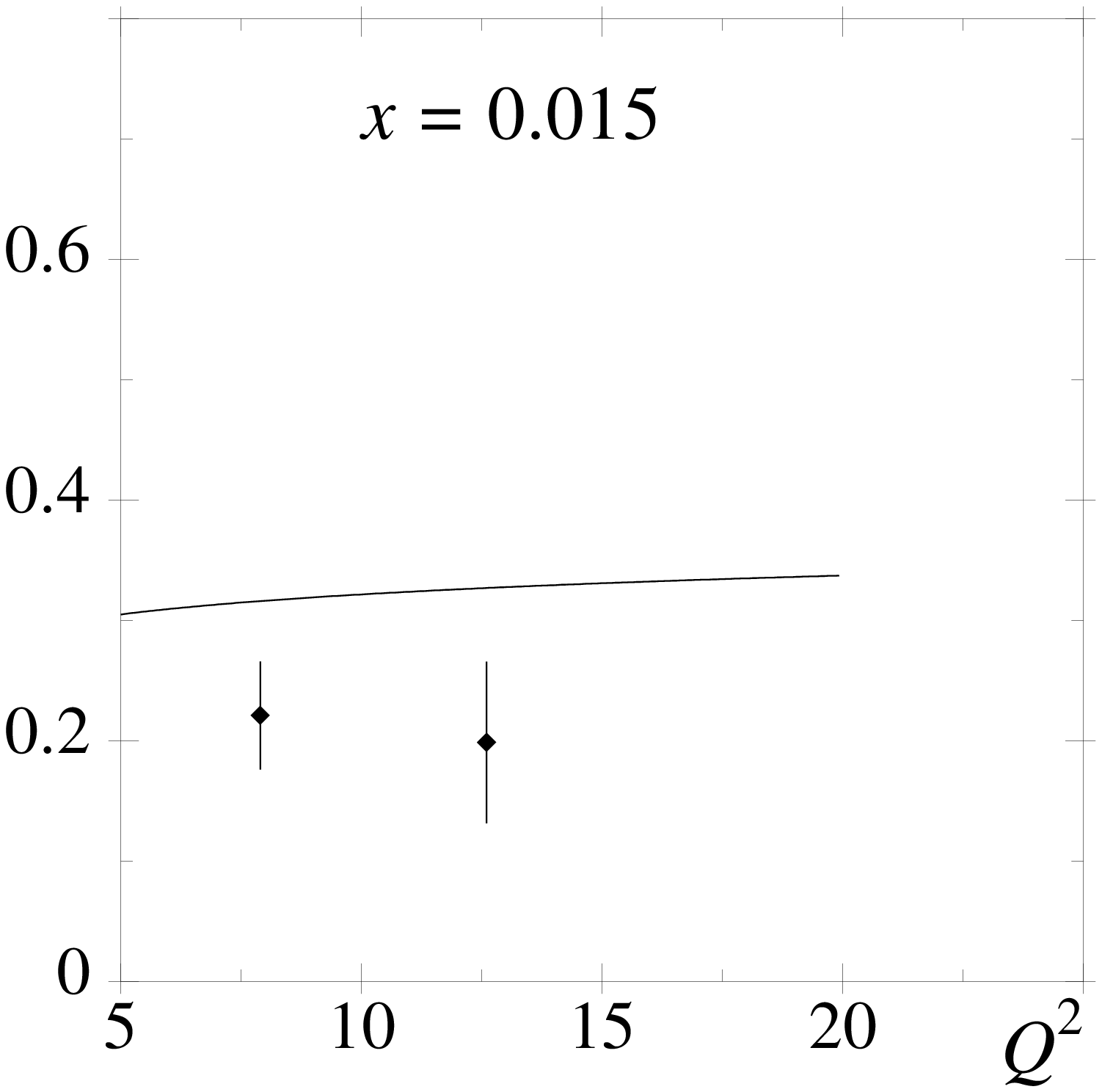}}
{\epsfxsize=3.3truecm\epsfbox{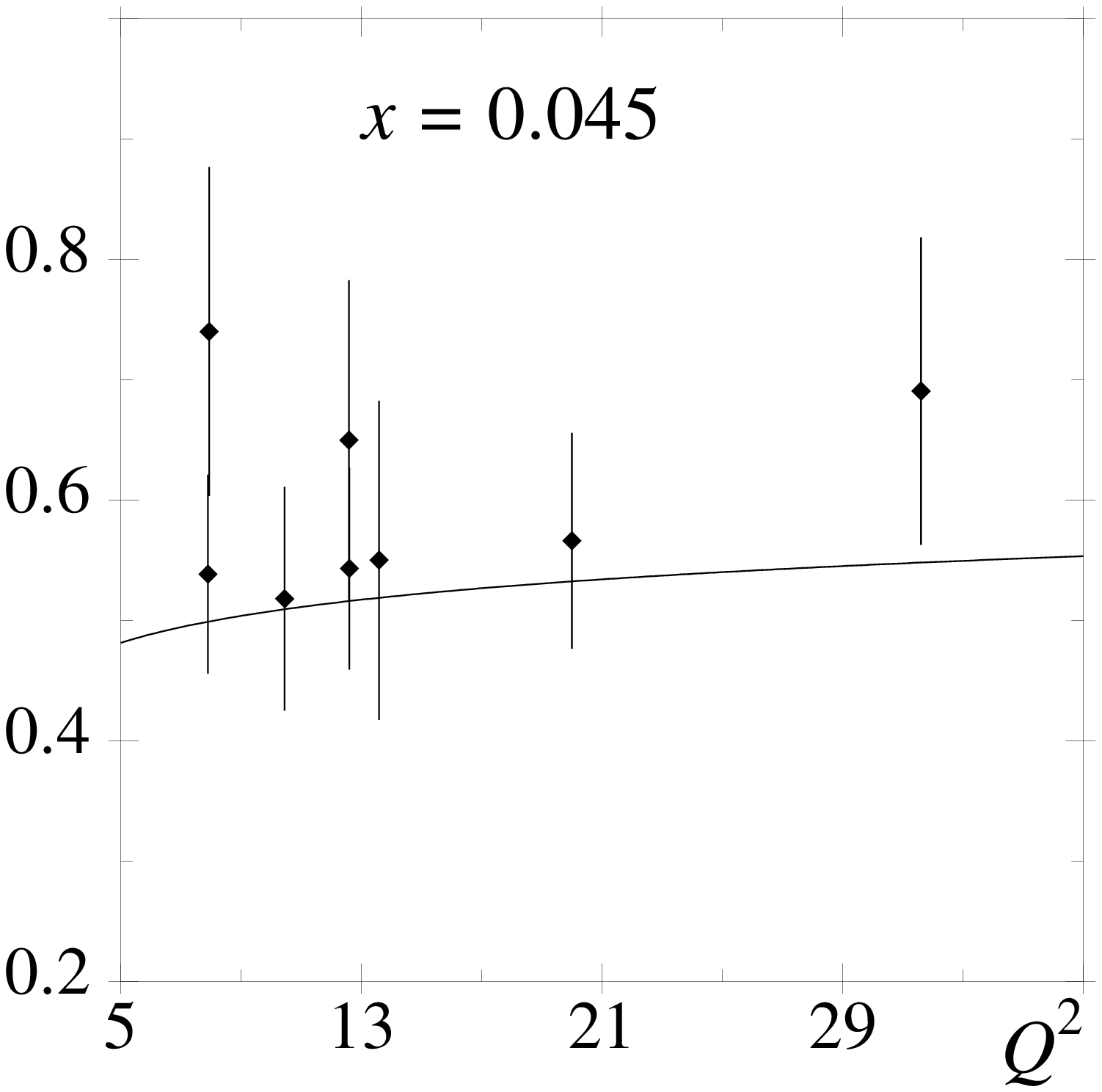}}}  }
\setbox0=\vbox{\hsize 6.7truecm \box12\hb{\petit
 Figure 2b. Comparison of predictions from 
Eq. (3.2c), Table V ($\Lambdav=0.14\;\gev$), with neutrino $xF_3$ data$^{[15]}$.\hb
\vskip.0cm}}
\line{\box1\hfil\box0}

 Finally, the fact that the  $\chi^2$ decreases with $\Lambdav$ past 
reasonable values   
is an indication that we are getting here an {\it effective} value for this 
parameter, which compensates for 
the large size of the NLO corrections.

We may consider fitting with the expanded version of the formula for $F_S$, i.e.,  
with
$$F_S(x,Q^2)\simeq B_S\left\{1+\frac{(c_S+q_S)\alpha_s}{4\pi}\right\}
\,\alpha_s^{-d_+}\,x^{-\lambda}(1-x)^{\nu_{\rm n.e.}(Q^2)},$$
and now
$$\eqalign{\nu_{\rm n.e.}(Q^2)=\nu_0+
\left\{\frac{B_S\left[1+(c_S+q_S)\alpha_s/4\pi\right]
\Gammav(1+\lambda)[16+3n_f]}{3n_f\alpha_s^{d_+}}\right\}^{1/(1-\lambda)};\cr
\nu_0=7.\kern13.5em\phantom{x}}$$
The fit {\it deteriorates} clearly; the $\chi^2$ is now of some 140, 
for 110-3 d.o.f. and $\Lambdav=0.10$. This shows that 
the exponentiated version of the formulas is to be preferred, as it probably sums at 
least part of the large NLO corrections. Because of this, we will 
henceforth use only the exponentiated version of the equations.

We next consider fits to the more recent Zeus data$^{[4]}$.
 We will make two choices: first, we fit the neutrino data, and 
all $ep$ points with $x\leq 0.01$ using the 
formula (2.10a). The results are given in Table VI. The 
chi-squared is reasonable, as is the value of $\Lambdav$. The values 
of all parameters 
are compatible with those found from 
 the fits to the H1 data. The second possibility is to extend the range to 
$x\leq 0.025$ and use Eq. (3.2), fixing $\Lambdav=0.135$.
 The results of the fit are shown in Fig. 3. We do not show the 
fit to the neutrino data, which does not differ substantially from that of Fig. 2b.
 The $\chi^2$/d.o.f. is now of 226.1/(120-3). 
This, as the $\chi^2$/d.o.f. reported for the fit 
of data with $x\leq 0.01$ in Table VI, are larger 
than their counterparts for the H1 data. A glance to Fig. 3 shows that part of the 
reason is the presence in the Zeus data 
of fluctuations. These are probably due to systematic errors not taken into account in 
the experimental analysis; they become 
important for very large $Q^2$. Thus, and although 
the Zeus data appear more precise than the H1 ones for the lower $Q^2$ range,\fonote{This 
is probably the reason why the value of $\Lambdav$ deduced 
from the H1 data is {\it less} realistic than 
that obtained fitting the set of Zeus.} 
the last one are more reliable at large $Q^2$. Nevertheless, and as 
noted in the comments to the fit to H1 data, it 
is also clear that the theoretical predictions present {\it systematic} 
deviations from experiment, very likely due to the extension of the 
first beyond their range of validity by use of a semiphenomenological 
expression which is not sufficiently flexible; see Sect. 5 for 
more discussion.
 Apart from this, the results are good and 
the parameters of the fits reasonably compatible. The value of $\Lambdav$ 
is closer to the accepted one.

\setbox0=\vbox{\hsize 14.7truecm \epsfxsize=14.6truecm\epsfbox{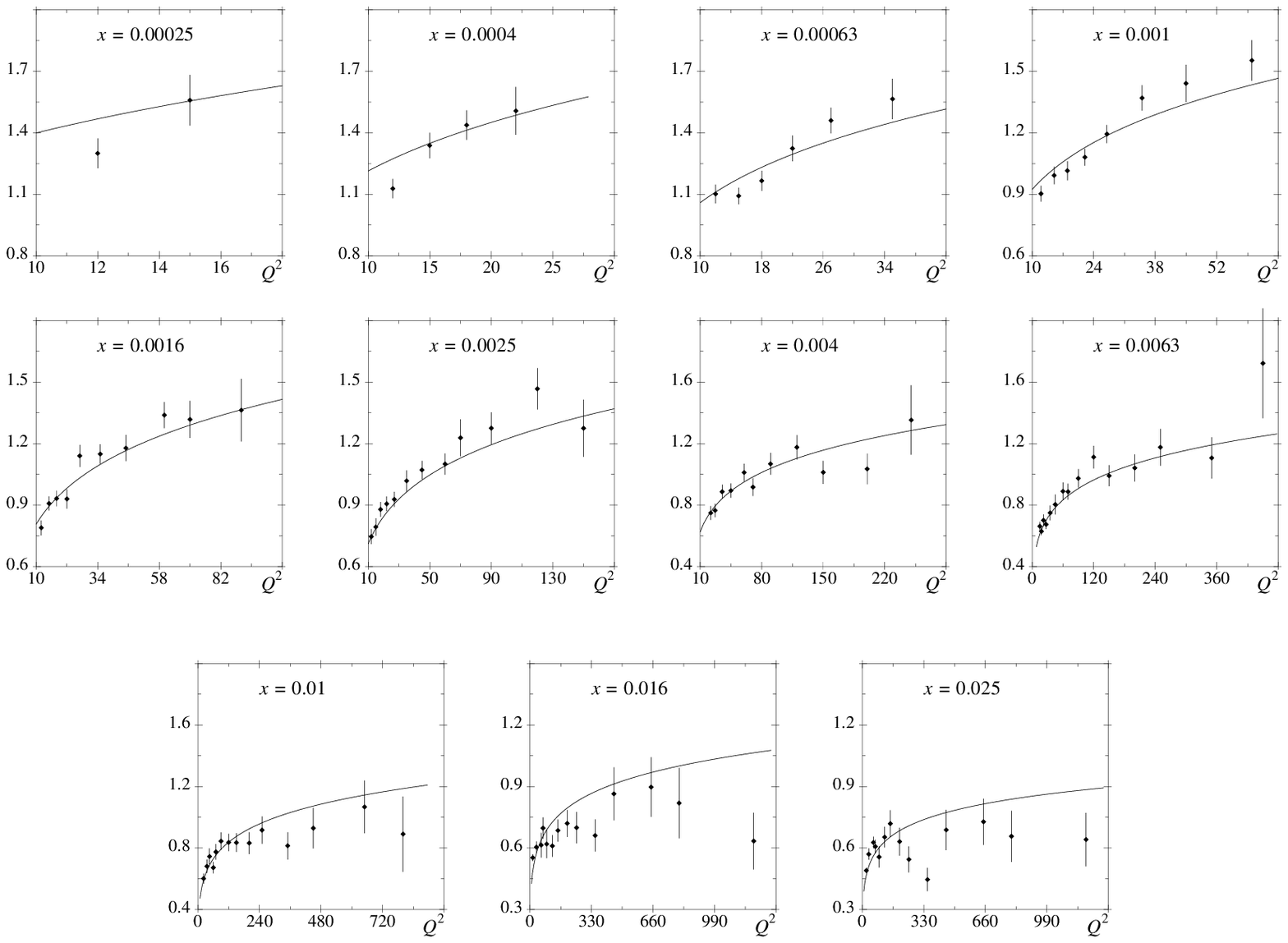}\hb{\petit
\phantom{x}\kern2cm Figure 3. Comparison of predictions for $F_2$ from 
Eqs. (3.2a, c), $\Lambdav=0.135$,
\hb\phantom{x}\kern2cm with Zeus $ep$ data
\vskip.2cm}}
\centerline{\box0}
 \vskip.2cm
\setbox0=\vbox{\hsize=14cm
{\petit 
\vskip.2cm
\centerline{{\bf Table VI}. Zeus plus $\nu$ data. $n_f=4;\,x\leq 0.01$.}
\centerrule{16em}
\vskip.1cm
 $$\matrix
{\Lambdav & \lambda & \langle e^2_q\rangle  B_S& \langle e^2_q\rangle  B_{NS}
&\chi^2/{\rm d.o.f.}\cr
& & & &\cr
0.135^{+0.075}_{-0.055}\,\gev&0.301\pm0.025&1.250^{+0.552}_{-0.400}\times 10^{-4}&0.3138\pm0.007
&\tfrac{126.4}{92-4}}$$
}
\vskip.1cm
}
\centerline{\boxit{\box0}} 

As a final check on the reliability and consistency of the 
fits we have fitted Zeus data with $x\leq 0.01$, {\it not} including the 
neutrino data. We get no definite minimum for $\Lambda$, 
only constrained by $\Lambda\lsim 0.2\,\gev$; but we obtain 
values of the remaining parameters compatible with those 
obtained {\it including} $xF_3$, in particular a very reasonable value 
for $B_{NS}$:
$$\langle e^2_q\rangle B_S=1.8\times 10^{-4}, \langle e^2_q\rangle B_{NS}=0.35,\lambda=0.329.$$
In this sense we may say that our analysis is sufficiently 
precise to {\it predict} the NS structure 
functions from $F_2$ only, 
and this in spite of the relative smallness of $F_{NS}$. It is however clear that, as 
already mentioned several times, systematic deviations occur, especially large for $x\gsim 0.01$  
(cf. \sect 5).
    
\vskip.2cm
{\bf 3.2. The gluon structure function.}
\vskip.2cm
We give here the parametrizations to  NLO for the 
gluon structure function that follow from our determination of the parameters in 
the previous subsection, for the full set of points 
corresponding to a set of parameters intermediate between those given in Table V for $F_2$,
$$\eqalign{F_G(x,Q^2)=B_G\left[1+\frac{-0.25 \alpha_s}{4\pi}\right]
\ee^{16.6\alpha_s/4\pi}\alpha_s(Q^2)^{-3.182}x^{-0.316}(1-x)^{\nu_G(Q^2)},\cr
{\nu_G(Q^2)}=5+
\left\{\frac{B_G\,[1-0.25 \alpha_s/4\pi]
\ee^{16.6\alpha_s/4\pi}
\Gammav(1+0.316)[16+3n_f]}{16\alpha_s^{3.182}}\right\}^{\tfrac{1}{1-0.316}},\cr 
B_G=25.2\times \langle e^2_q\rangle B_S,\;\langle e^2_q\rangle B_S=1.226\times 10^{-4},\cr
{\rm NLO},\; x\leq 3.2\times10^{-2},\;12\;\gev^2\leq Q^2\leq 1\,200\;\gev^2;\cr
\,n_f=4;\,\Lambdav(2\,{\rm loop},n_f=4)=0.14\,\gev.}\eqno (3.5)$$
The corresponding graphs are shown in Fig. 4, where we give both LO and NLO predictions, the LO 
calculation with values of parameters from Table I.

There are unfortunately no direct measurements of $F_G$ with 
which to compare our calculations. {\it Indirect} estimates 
were made by the H1 collaboration,
 by fitting $F_2,\,F_G$ with an exact coupled QCD evolution. The comparison 
with our calculations to LO may be found in ref. 1; the agreement is reasonable, and indeed our 
estimates are more precise than the DGLAP calculation,
 afflicted by large extrapolation errors.

More information on $F_G$ is obtained from the cross-section $\gamma p\rightarrow J/\psi p$, 
to be discussed in \sect 5.3. 

\setbox3=\vbox{\hsize 14.7truecm \epsfxsize14.6truecm\epsfbox{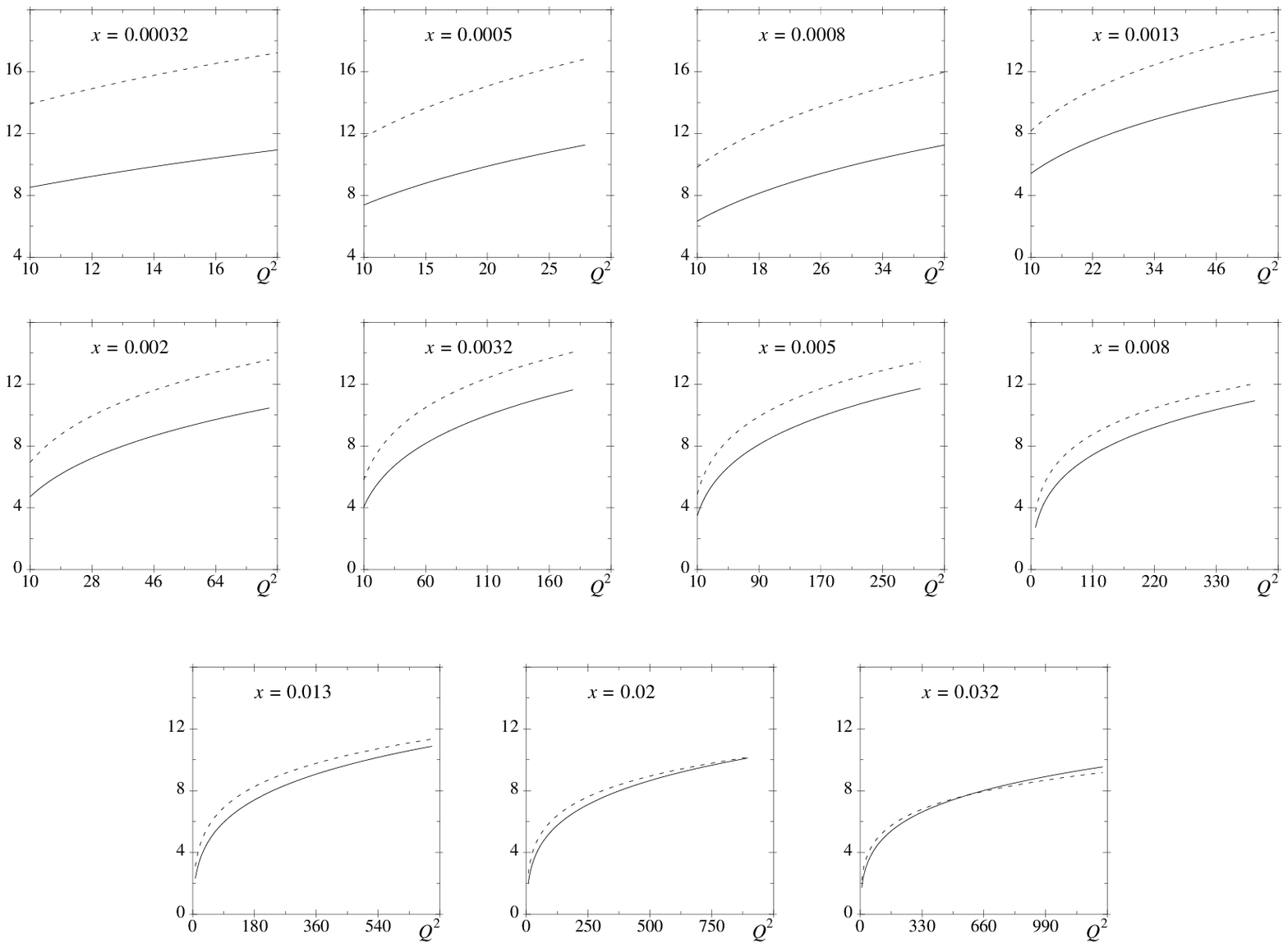}\hb{\petit
\phantom{x}\kern2cm Figure 4. The gluon structure function $F_G$
 to LO (broken line), \hb
\phantom{x}\kern2cm Eq. (3.3), and the optimum NLO one (continuous line), Eq. (3.5).
 \hb
\vskip.2cm}}
\centerline{\box3}
 \vskip.2cm
{\bf 3.3. Predictions for the longitudinal structure function.}
\vskip.2cm
\setbox4=\vbox{\hsize=7.5truecm 
 NLO, $O(\alpha_s^2)$ corrections to the longitudinal
 structure function are unfortunately {\it very} large; 
not because of the direct corrections, but due to corrections generated indirectly via 
the large NLO corrections to $F_2$. Indeed, the 
value of $R'$ is reduced by more than a half from LO to NLO. We give in 
Fig. 5 a plot of LO and NLO calculations. Using Eq. (2.14), and the 
parameters $\Lambdav=0.20,\;\lambda=0.38$ (Table I{\it a}) we get the  LO result, $R^{(0)}$;
 and with Eqs. (2.15), (2.16) and the figures  
 $\Lambdav=0.10,\;0.20\;{\rm and}\;\lambda=0.324$ (cf. Table IV{\it a})
 we find $R'$ (NLO).  Also depicted are a few 
representative data. Note that the dependence of the NLO value 
of $R$ on $\Lambdav$ is very slight, due to cancellation of 
various effects. Thus, the lines corresponding to $\Lambdav=0.10,\;0.20$ 
in Fig. 5 fall almost one on top of the other.}
\setbox1=\vbox{\epsfxsize=7.0truecm \epsfbox{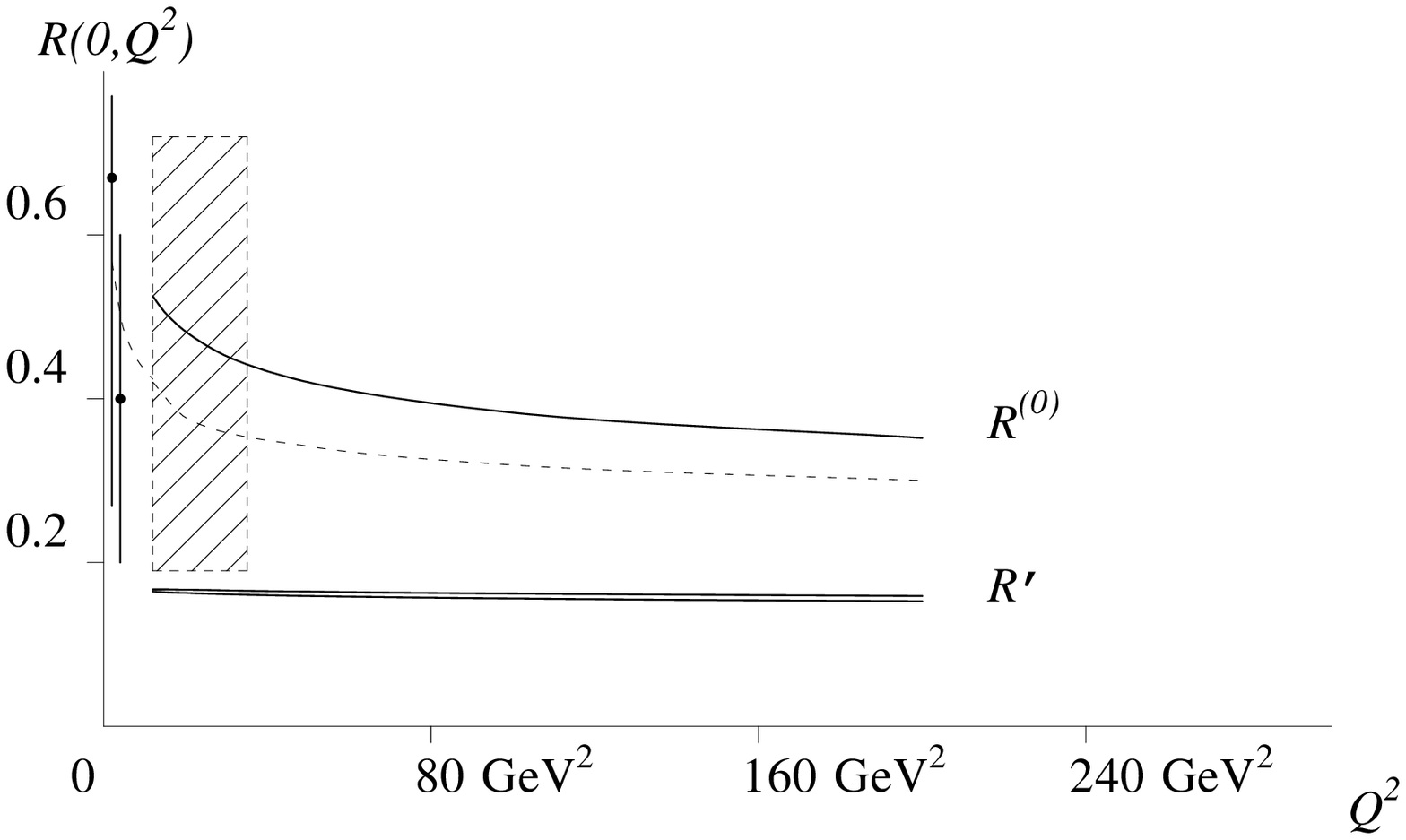}}
\setbox3=\vbox{\hsize6.2truecm\petit \noindent Figure 5. Predictions for $R'(0,Q^2)$ to
 LO and NLO. Hatched box: 
preliminary result from H1$^{[17]}$. Dots: data from ref. 18 (actually, at 
$x\sim0.05$). Discontinuous line: intermediate calculation, see main text.\hb\vskip.2cm}
\setbox0=\vbox{\hsize=7.2truecm\box1\vskip.2cm\rightline{\box3}}
\line{\box4\hfil\box0}

 To get a further indication on the meaning of the results,
 we have also calculated $R'$ from the effective fit
 at low energy of ref. 1, with $\lambda =0.324$, 
and $\alpha_s$ to two loops, but without other NLO corrections, 
taking $n_f=3$ below $Q^2=12\;\gev^2$, and $n_f=4$ above. This is
 the intermediate, dashed curve in Fig. 5. Clearly, one would expect 
that the real $R'$ would somehow interpolate between this, at low momentum, and the full
 NLO curve, for very large $Q^2$.
The predictions should be checked against experiment when, and if, measurements independent 
of those of $F_2$ are performed 
at HERA. We have given the predictions for $x=0$; the figures
 would not change much provided $x\leq 10^{-2}$.
\vskip.4cm
\centerline{4. {\bf THE SOFT-POMERON DOMINATED MODEL}}
\vskip0.5cm
As remarked in the Introduction, the results derived in the previous sections assume that 
the singlet structure functions are dominated, at small $x$, by the singularities of 
the matrix elements of the quark and gluon operators. We may instead hypothesize that 
these singularities lie to the left of $n=1$, and then the small $x$ behaviour is 
controlled by the singularities of the Wilson coefficients. Specifically, this 
occurs if one assumes that, for all $Q^2$ below a certain $Q_0^2$ 
of the order of a typical hadronic scale (say $\sim$ GeV), cross sections behave according to 
a standard soft-Pomeron dominance,
$$\sigma_{\rm tot}(Q_0^2,s)\simeq {\rm Constant},$$
 for $Q_0^2\lsim 1\;\gev^2$.

We can then get the structure functions for small $x$ by evolving with QCD the expressions 
corresponding to this,$^{[9,10,11]}$
$$F_i(x, Q_0^2)\simeqsub_{x\rightarrow 0}c_i,\;i=S,\,G.\eqno (4.1)$$
We will present a sketchy derivation of the 
resulting formulas, to NLO. This is of interest because we use the moments method, instead 
of the Altarelli-Parisi equations employed in ref. 11, so we have 
a nontrivial check of the calculation there.
\vskip1cm

{\bf 4.1. Theoretical calculations: $F_S$ and $F_G$}
\vskip.2cm
Our starting point is the following relation, proved in 
 ref. 6 to NLO,
$$\alpha_s(Q^2)^{\hat{\bf D}}(1+a{\bf \Gamma}){\bf S}^{-1}(1-a{\bf C}^{(1)})\ybf{\mu}(n, Q^2)
\equiv {\bf b}={\rm independent\;of}\;Q^2.\eqno (4.2)$$
Here $a=\alpha_s(Q^2)/4\pi$, ${\bf C}^{(1)}$ is 
the matrix of NLO corrections to the Wilson coefficients\fonote{Defined as in ref. 6,
 which fixes the arbitrariness in $F_G$.} and 
$$\alpha_s^{\hat{\bf D}}=\left(\matrix{\alpha_s^{d_+}&0\cr0&\alpha_s^{d_-}}\right),\;
{\bf\Gamma}=-\frac{1}{2\beta_0}
\left(\matrix{\bar{\gamma}_{11}(n)+2\beta_1d_+(n)&
\dfrac{\bar{\gamma}_{12}(n)}{d_+(n)-d_-(n)+1}\cr
\dfrac{\bar{\gamma}_{21}(n)}{d_-(n)-d_+(n)+1}&
\bar{\gamma}_{22}(n)+2\beta_1d_-(n)}\right)
\eqno (4.3)$$
with ${\bf S},\,\bar{\gamma}$ defined before;
explicit expressions for these quantities may be found in the Appendix.  Here we only give the 
values, in the limit $n\rightarrow 1$, of 
those of interest for us now. We have:
$$\eqalign{d_+(n)\simeq\dfrac{d_0}{4(n-1)}-d_1,\;d_-(n)\simeq -\dfrac{16n_f}{27\beta_0},\cr
d_0=\dfrac{48}{\beta_0},\;d_1=\dfrac{11+\tfrac{2}{27}n_f}{\beta_0};\cr
{\bf S}(n)\simeq\left(\matrix{1&-\dfrac{n_f(n-1)}{9}\cr
\dfrac{9}{n_f(n-1)}&\dfrac{4n_f(n-1)}{81}}\right),}\eqno (4.4{\rm a})$$
 and, defining $\gamma_{ij}(n)\simeq r_{ij}/(n-1)$,
$$\eqalign{r_{12}=-\tfrac{320}{9}n_fT_FC_A,\cr
r_{22}=\tfrac{368}{9}n_fT_FC_A-\tfrac{32}{3}n_fT_FC_F.}\eqno (4.4{\rm b})$$
Finally,
$$\eqalign{C^{(1)}_{11}(n)\simeq C_F[\pi^2-\tfrac{17}{2}-4\zeta(3)](n-1),\;
C^{(1)}_{12}(n)\simeq \tfrac{4}{3}n_fT_F\cr
C^{(1)}_{21}(n)\simeq \dfrac{16T_F}{3(n-1)},\;C^{(1)}_{22}(n)\simeq \dfrac{12T_F}{n-1}.}
\eqno (4.4{\rm c})$$
In the soft Pomeron hypothesis the behaviour of $F_{S,G}$ as $x\rightarrow 0$ is, as 
discussed,  
dominated by the singularities of the ${\bf C}^{(1)}(n),\;
\ybf{\gamma}^{(0)}(n),\;\ybf{\gamma}^{(1)}(n)$ as $n\rightarrow 1$, which in turn 
give those of the $\ybf{\mu}(n)$. From (4.2 - 4) one easily finds,
$$\eqalign{\ybf{\mu}(n,Q^2)\simeqsub_{n\rightarrow 1}b_1\alpha_s^{-d_+(n)}
\left(\matrix{1+\dfrac{ak_1}{n-1}\cr
\dfrac{9}{n_f(n-1)}\left\{1+\dfrac{ak}{n-1}\right\}}\right),\cr
k=12T_F+\dfrac{1}{2\beta_0}\left(\tfrac{4}{9}r_{12}+r_{22}+\dfrac{24\beta_1}{\beta_0}\right),
\;k_1=k-\dfrac{3r_{12}}{8n_f}.}
\eqno (4.4{\rm d})$$
Note that, because for $n\rightarrow 1$, $d_+(n)\gg d_-(n)$, only 
the term in $\alpha_s^{-d_+(n)}$ (and not that in $\alpha_s^{-d_-(n)}$) 
contributes.

Next we evaluate $b_1$ in terms of $F_i(x,Q_0^2)$, assumed to behave as 
in (4.1) so that
$$\mu_i(n,Q_0^2)\simeqsub_{n\rightarrow 1}c_i\dfrac{1}{n-1}.\eqno (4.5)$$
This is accomplished using again (4.2) for $Q=Q_0$, 
profiting from the independence of $b_1$ on $Q^2$. The computation is 
 straightforward and we find
$$\eqalign{b_1=\alpha_s(Q^2_0)^{d_+(n)}\left\{1
-\dfrac{\alpha_s(Q^2_0)k}{4\pi(n-1)}\right\}c_0,\cr
c_0=n_f\left[\tfrac{4}{81}c_1+\tfrac{1}{9}c_2\right].}\eqno (4.6)$$
 Plugging the result into (4.4) we get 
$$\eqalign{\ybf{\mu}(n,Q^2)\simeqsub_{n\rightarrow 1}c_0\ee^{\tau d_0/4(n-1)-\tau d_1}\cr
\times\left(\matrix{1+\left[k-\dfrac{3r_{12}}{8n_f}\right]\dfrac{\alpha_s(Q^2)}{4\pi(n-1)}-
k\dfrac{\alpha_s(Q^2_0)}{4\pi(n-1)}\cr
\dfrac{9}{n_f(n-1)}\left\{1+k\dfrac{\alpha_s(Q^2)}{4\pi(n-1)}-
k\dfrac{\alpha_s(Q^2_0)}{4\pi(n-1)}\right\}}\right),\cr
\tau=\log\dfrac{\alpha_s(Q^2_0)}{\alpha_s(Q^2)},}\eqno (4.7)$$
with $d_0,\,d_1$ as in (4.4a).
We then invert the Mellin transform. Generally, if
$$\mu(n)=\int^1_0\dd x\;x^{n-1}F(x),\eqno (4.8{\rm a})$$
and
$$\mu(n)\simeq\dfrac{1}{(n-1)^{\nu}}\ee^{d_0 \tau/4(n-1)}\eqno (4.8{\rm b})$$
then 
$$F(x)\simeqsub_{x\rightarrow 0\atop \tau\rightarrow \infty}
\xi^{\sigma}f_{\sigma}(\tau)\ee^{\sqrt{d_0\tau\xi}}\eqno (4.9)$$
where\fonote{From (4.8 to 10) it thus follows that powers of $n-1$ correspond 
to powers of $\sqrt{|\log x|}$. Therefore, we cannot, unless a more definite 
assumption is made about the behaviour of the structure functions at 
$Q^2_0$, give results more precise than 
terms of relative order $1/\sqrt{|\log x|}$.}
$$\xi=\log x^{-1},\;\sigma=\tfrac{1}{2}\nu-\tfrac{3}{4},
\;f_{\sigma}(\tau)=\dfrac{4^{\sigma}}{\pi^{\frac{1}{2}}}(d_0\tau)^{-\sigma-\frac{1}{2}}.
\eqno (4.10)$$
The proof is elementary and is obtained by substituting (4.9) in (4.8a)  
and integrating. Thus we get the final result,
$$\eqalign{F_S(x,Q^2)\simeqsub_{x\rightarrow 0\atop Q^2\gg Q_0^2}
c_0\xi^{-1}\left(\dfrac{d_0\tau\xi}{64\pi^2}\right)^{\frac{1}{4}}\exp\left[\sqrt{d_0\tau\xi}-d_1\tau\right]\cr
\times\left\{1+2\sqrt{\dfrac{\xi}{ d_0\tau}}\left[k_1\;
\dfrac{\alpha_s(Q^2)}{4\pi}-
k(1+\delta)\;\dfrac{\alpha_s(Q^2_0)}{4\pi}\right]\right\},}
\eqno (4.11{\rm a}),$$
$$\eqalign{F_G(x,Q^2)\simeqsub_{x\rightarrow 0\atop Q^2\gg Q_0^2}\dfrac{9c_0}{n_f}
\dfrac{1}{(4\pi^2 d_0 \tau\xi)^{\frac{1}{4}}}\exp\left[\sqrt{d_0\tau\xi}-d_1\tau\right]\cr
\times\left\{1+2\sqrt{\dfrac{\xi}{d_0\tau}}
\left[k\;\dfrac{\alpha_s(Q^2)}{4\pi}-k(1+\delta)\;\dfrac{\alpha_s(Q^2_0)}{4\pi}\right]\right\}.}
\eqno (4.11{\rm b})$$
Here $k,\,k_1$ are given in (4.4d) with the $r_{ij}$ of (4.4b). We have added an 
arbitrary  factor $(1+\delta)$ 
for reasons that will be clear later. In the soft Pomeron model, one of course has 
$\delta=0$. Numerically,
$$\relax \eqalign{k_1=k-3r_{12}/8n_f\simeq 42.19\cr
k\simeq 22.19\; ({\rm both\;for}\;n_f=4).}\eqno (4.11{\rm c})$$

Eq. (4.11a) may be compared with the calculation of Ball and Forte.$^{[11]}$ We agree in the LO 
term, and in the coefficient of $\alpha_s(Q^2)/4\pi$ in the NLO term, but disagree in the 
coefficient of the $\alpha_s(Q^2_0)/4\pi$ term. This is not of great
 moment\fonote{Nevertheless an independent calculation 
that resolved the discrepancy would be welcome.} since the numerical 
difference is slight, 22.19 {\it vs} 16.19. Eq. (4.11b) is given 
here for the first time.

NLO corrections are {\it very} large. Indeed, for fixed
 $Q^2,\,x\rightarrow 0$, the NLO correction 
overwhelms the LO part. This, together with the problem posed by 
the BFKL-Florentine terms$^{[19,20]}$
$$x^{-\omega_0\alpha_s(\nu^2)};\;\omega_0={\rm Constant}$$
will be discussed in Sect. 5.
\vskip1cm
{\bf 4.2. Theoretical calculations: longitudinal structure function.}
\vskip.2cm
We define as before (Sect. 2.3) $R'\simeq F_L/F_S$. Because, 
in the soft Pomeron dominated model, the 
contribution to $F_L$ of $F_S$ is subleading with 
respect to that of $F_G$ in the $x\rightarrow 0$
 limit, it follows that we may, to errors of relative size $\sqrt{1/|\log x|}$, neglect 
the contribution of $F_S$ to $F_L$. 
For completeness, however, we will give the formula including   
 this contribution of $F_S$. We then have an equation similar to (2.13),
$$F_L(x,Q^2)\simeq F_G(x,Q^2)\int_0^1\dd y\;C_G^L(y,Q^2)+
F_S(x,Q^2)\int_0^1\dd y\;C_S^L(y,Q^2),\eqno (4.12)$$
and the $C^L$ are as in (2.12b).

We let
$$\eqalign{I_g(x)=T_Fn_f[C_FI_{G1}(0)+C_AI_{G2}(0;x)]
\simeq -17.62-\dfrac{32n_fT_FC_A}{9}\log x^{-1},\cr
I_q(x)=C_F(C_A-2C_F)I_{S1}(0)+C_F^2I_{S2}(0)+C_FT_Fn_fI_{S3}(0)
+C_FT_Fn_fI_{PS}(0;x),}\eqno (4.13)$$
where 
$$\int^1_x\dd yc^{(1)L}_G(y)\equiv n_fT_F[C_FI_{G1}(0)+C_AI_{G2}(0;x)],$$
and similar expressions for the $I_S,\,I_{PS}$. We then obtain the NLO expression 
for $R'$, 
$$\eqalign{R'(x,Q^2)\simeqsub_{x\rightarrow 0\atop Q^2\gg Q^2_0}
\Bigg\{2C_F+I_q\dfrac{\alpha_s(Q^2)}{4\pi}\cr
+48T_F\sqrt\dfrac{\xi}{d_0\tau}
\Big[1+\big(-\tfrac{80}{3}T_FC_A\sqrt\dfrac{\xi}{d_0\tau}+
\tfrac{3}{8}I_g\big)\dfrac{\alpha_s(Q^2)}{4\pi}\Big]\Bigg\}\dfrac{\alpha_s(Q^2)}{4\pi}.}
\eqno (4.14)$$

Because the NLO corrections are so large, we will use, instead 
of (4.14), a nonexpanded version for comparison with experiment. Removing also the NLO 
contribution of $F_S$ the formula to be 
employed for numerical calculations is then,
$$\eqalign{R'(x,Q^2)\simeqsub_{x\rightarrow 0\atop Q^2\gg Q^2_0}
\Bigg[48T_F\sqrt{\dfrac{\xi}{d_0\tau}}+2C_F\Bigg]
\dfrac{1+\left(2k\sqrt{\xi/d_0\tau}\right)\alpha_s(Q^2)/4\pi}
{1+\left(2k_1\sqrt{\xi/d_0\tau}\right)\alpha_s(Q^2)/4\pi}\cr
\times\left\{1+\tfrac{3}{8}I_g(x)\,\dfrac{\alpha_s(Q^2)}{4\pi}\right\}\dfrac{\alpha_s(Q^2)}{4\pi}.
}
\eqno (4.15)$$

\vskip1cm
{\bf 4.3. Comparison with experiment.}
\vskip.2cm
For the soft Pomeron dominated model a very peculiar phenomenon occurs:
 the LO expressions produce fits {\it better} than the NLO ones. What is more, and
 unlike in the hard singularity case where we could blame the discrepancy on the 
large $x$ points, here it is uniformly distributed. The strategy for 
comparison with experiment should be different now. First of all, we will not 
include a term like $(1-x)^{\nu}$ connected with the saturation of the  
momentum sum rule since it is now very small, and would arrange nothing. Secondly, we 
give parameters for the LO fit for the restricted ($x\leq 0.01$) range, and   
we give results of the NLO calculation both for 
the restricted ($x\leq 0.01$) and full ranges. These we will discuss in greater detail.

As stated we begin a LO calculation, 
fixing $\Lambdav(1\;{\rm loop}\; n_f=4)=0.20\;\gev$, and taking for 
definiteness the H1 data, plus the neutrino 
data for stability. We find the results of Table VII{\it a}.
\vskip.2cm
\setbox0=\vbox{\hsize=14cm
{\petit
\vskip.2cm
\centerline{{\bf Table VII}{\it a}. LO calculation; $x\leq 0.01$, H1 plus $\nu$ data.}
\vskip.1cm
\centerrule{16em}
$$\matrix
{\langle e_q^2\rangle c_0&\langle e_q^2\rangle B_{NS}&Q_0^2&\chi^2/{\rm d.o.f.}\cr
 & & &\cr
0.12&0.44&0.46\;\gev^2&\tfrac{48.2}{68-3}\cr
}$$
}
\vskip.1cm
}
\centerline{\boxit{\box0}} 

We consider next the NLO calculation which we split into two parts: restricted 
range, and full range. For the first we give the results of the calculation in Table VII{\it b}. Only the 
H1 data are considered, for comparison with Table VII{\it a}. We {\it do} include 
neutrino data to force reasonable values for $B_{NS}$.
\vskip.2cm
\setbox0=\vbox{\hsize=14cm
{\petit
\vskip.2cm
\centerline{{\bf Table VII}{\it b}. NLO calculation; $x\leq 0.01$, H1 plus $\nu$ data.}
\vskip.1cm
\centerrule{16em}
$$\matrix
{\Lambdav&\langle e_q^2\rangle c_0&\langle e_q^2\rangle B_{NS}&Q_0^2&\chi^2/{\rm d.o.f.}\cr
 & & &\cr
0.160_{-60}^{+70}\gev&0.292{\pm 0.001}&0.326\pm{0.020}&0.86^{+0.40}_{-34}\;\gev^2
&\tfrac{74.3}{68-4}\cr
}$$
}
\vskip.1cm}
\centerline{\boxit{\box0}}
As mentioned, the chi-squared has clearly 
deteriorated,\fonote{For Zeus data we would have had a much worse figure,
 $\chi^2$/d.o.f.=$\tfrac{180.5}{92-4}$, but a slightly 
better $\Lambdav=0.20$. The other parameters do not change substantially.}
 although the values of $Q_0^2,\,B_{NS}$ are 
more realistic now. The value of $\Lambdav$, also fitted, is reasonable. 

For the full range we give the results of the fits to both H1 and Zeus data  
in Tables VIII. For the H1 set there is no reasonable minimum for $\Lambdav$; 
for Zeus the optimum is for $\Lambdav=0.165\,\gev$. Thus we fix this value for 
both sets of data.
\vskip.2cm
\setbox0=\vbox{\hsize=14cm
{\petit
\vskip.2cm
\centerline{{\bf Table VIII}{\it a}. NLO calculation; Zeus plus $\nu$ data, $x\leq 0.025,\,\Lambda=0.165\,\gev$}
\vskip.1cm
\centerrule{16em}
$$\matrix
{\langle e_q^2\rangle c_0&\langle e_q^2\rangle B_{NS}&Q_0^2&\chi^2/{\rm d.o.f.}\cr
 & & &\cr
0.282&0.240&0.90\;\gev^2&\tfrac{273.2}{120-4}\cr
}$$   
\vskip.1cm
\hrule
\vskip.2cm
\centerline{{\bf Table VIII}{\it b}. NLO calculation; H1 plus $\nu$ data, $x\leq 0.032,\,\Lambda=0.165\,\gev$}
\vskip.1cm
\centerrule{16em}
$$\matrix
{\langle e_q^2\rangle c_0&\langle e_q^2\rangle B_{NS}&Q_0^2&\chi^2/{\rm d.o.f.}\cr
 & & &\cr
0.265&0.246&0.70\;\gev^2&\tfrac{190.6}{110-4}\cr
}$$    
}
\vskip.1cm
}
\centerline{\boxit{\box0}}
 Our results are 
consistent among themselves, and with an
 existing NLO calculation, based on H1 data$^{[21]}$; the comparison with the Zeus data 
for $x\leq 0.025$ is shown in Fig. 6.

\setbox2=\vbox{\hsize16.1truecm \centerline{{\epsfxsize=16.truecm\epsfbox{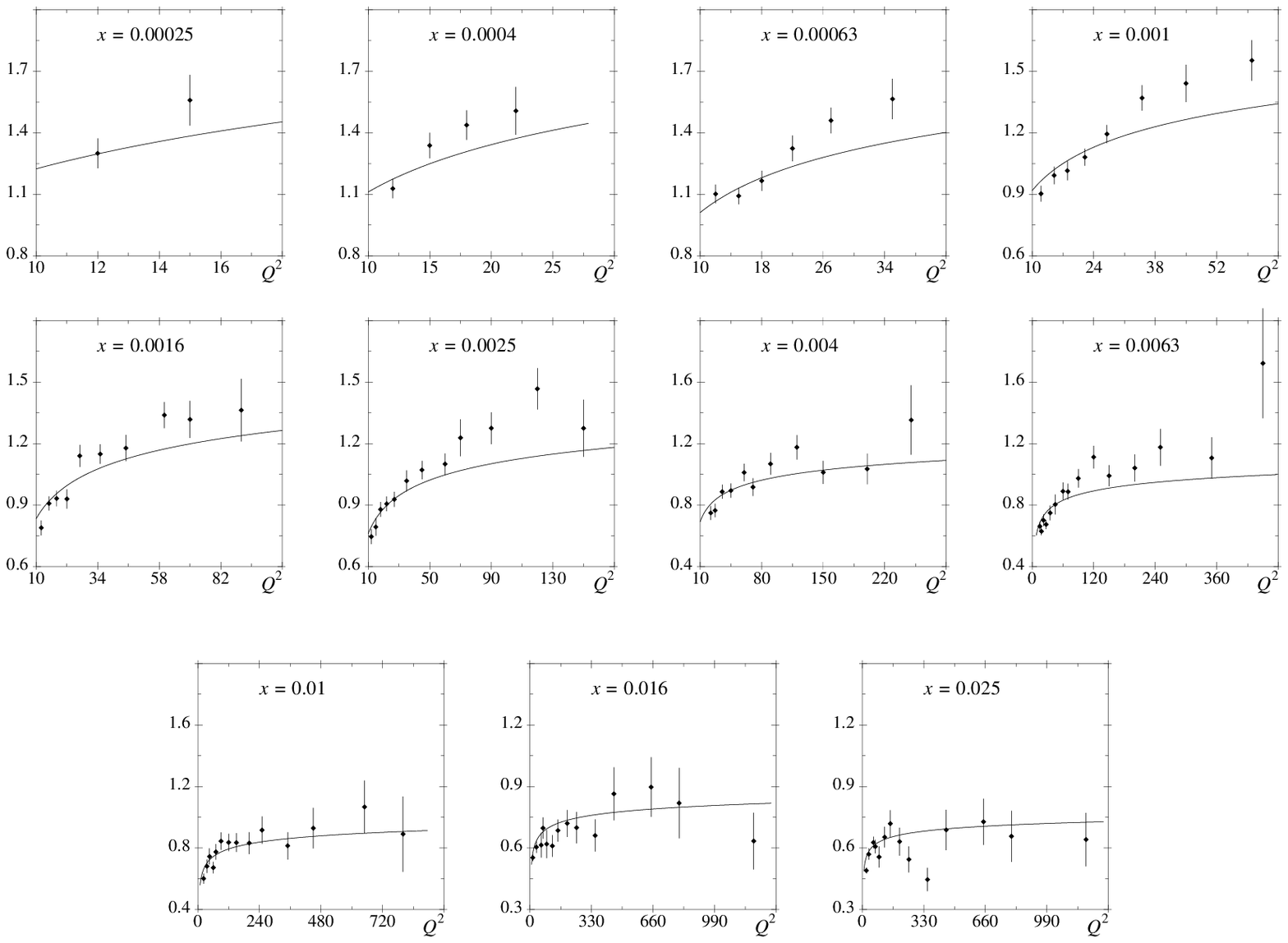}}}}
\setbox3=\vbox{\hsize 16.2truecm \box2\hb{\petit
\phantom{x}\kern2cm Figure 6. Predictions for the structure function $F_2$
 to NLO \hb
\phantom{x}\kern2cm in the soft-Pomeron model, and Zeus data.
\vskip.1cm}}
\centerline{\box3}

The results may be confronted with the ones obtained if {\it not} including 
the NLO correction: we would have obtained, for the generally 
preferred value $\Lambdav=0.23\,\gev$, 
$$\eqalign{\chi^2/{\rm d.o.f.}=\frac{142.1}{120-3}\;({\rm Zeus})\cr
\chi^2/{\rm d.o.f.}=\frac{90.0}{110-3}\;({\rm H1}).}$$
The situation is somewhat unpleasant. To make it worse, we mention that, if we 
delete the term in $\alpha_s(Q^2_0)$ in Eq. (4.11a), by simply 
putting $\delta=-1$ there, the quality of the fit {\it improves} 
substantially: to a chi-squared/d.o.f. of $\frac{141.4}{120}$ for 
Zeus data, and $\frac{78.1}{110-3}$ for H1, with $\Lambdav=0.23\,\gev$.

It is difficult to draw a clear-cut conclusion from this. At any rate, in all cases 
the fits are comparable in quality to those 
obtained with the hard singularity hypothesis, 
and reasonably good; more discussion will be given in Sect. 5. 

\setbox4=\vbox{\hsize=7.5cm For the longitudinal function, the predictions and 
comparison with experiment are depicted in Fig. 7; 
both the LO prediction based on the parameters of Table VII, and 
the NLO ones using the figures from Table VIIIa. Like in the 
hard singularity case, and for the same reason (large size of NLO corrections to $F_S$) there 
is a dramatic decrease between LO and NLO predictions,
 particularly for ``large" values of $x$, and for {\it very} small ones.
 NLO results, depicted for 
various values of $x$ in the figure are
 below the data.
 One cannot, nevertheless, consider the disagreement with experiment
 to be serious given the errors  both of it and of the theory. Perhaps more 
serious is the problem that the NLO 
corrections also here overwhelm the LO piece for $x\rightarrow 0$. }
\setbox1=\vbox{\epsfxsize=7.truecm \epsfbox{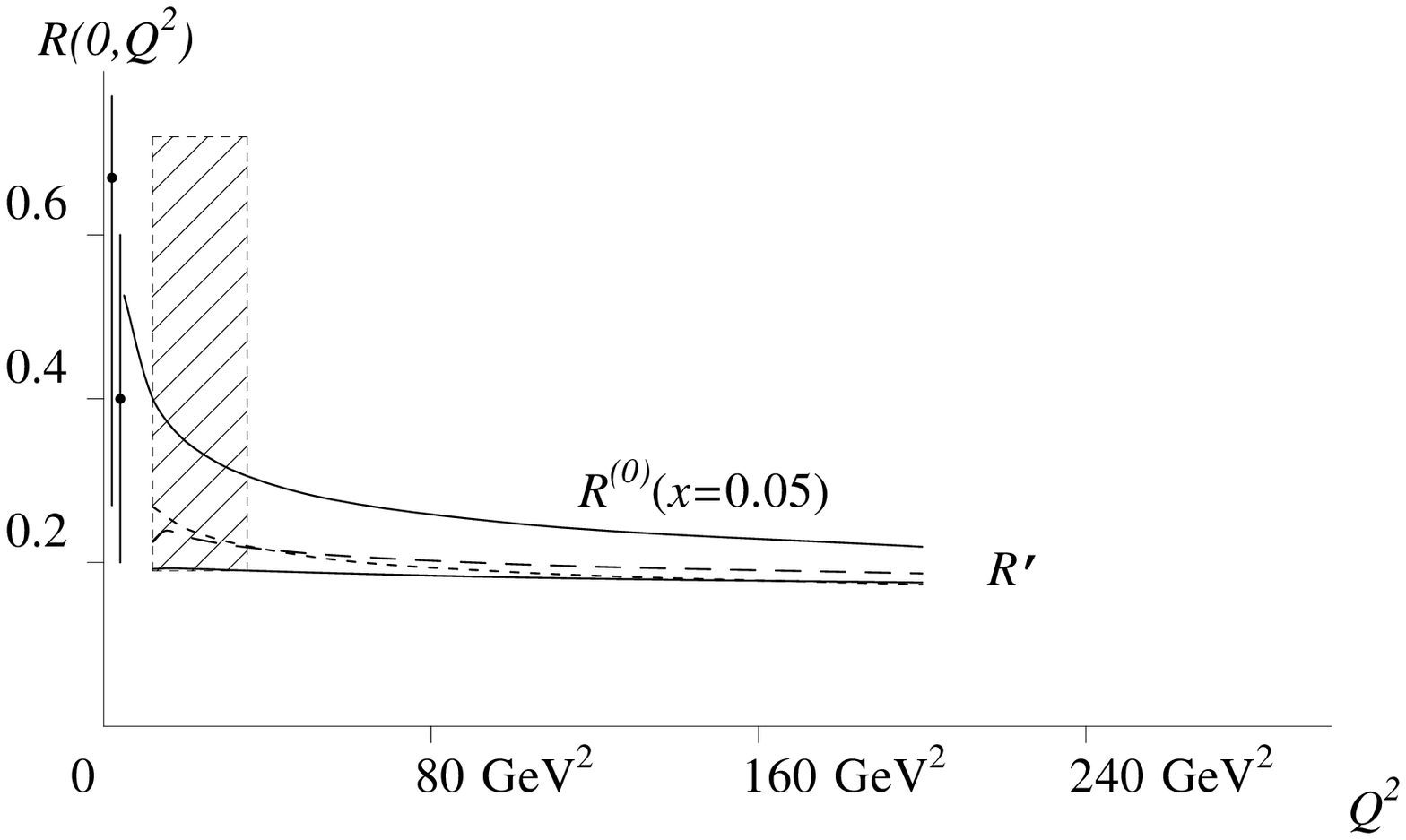}}
\setbox3=\vbox{\hsize6.1truecm\petit \noindent Figure 7. Predictions for $R'(x,Q^2)$ to
 LO and NLO, soft Pomeron model. $R'$:
  dotted, $x=0.05$;  dashed line: $x=10^{-3}$; continuous 
line: $x=10^{-4}$. Hatched box: ref. 17 ($x\sim10^{-3}$). Dots: data from ref. 18 at 
$x\sim0.05$.\hb\vskip.2cm}
\setbox0=\vbox{\hsize=7.5truecm\box1\vskip.2cm\rightline{\box3}}
\line{\box4\hfil\box0}
  
\vskip.4cm
\centerline{{5. \bf HARD PLUS SOFT SINGULARITIES. LARGE $Q^2$.}}
\vskip0.5cm
{\bf 5.1. Discussion.}
\vskip.2cm
It may appear strange that two mutually contradictory hypotheses, 
leading to so apparently different behaviours as the soft and hard Pomeron ones, produce
 both results in fair agreement with the data. The reason, however, is 
not difficult to find: both behaviours solve the QCD evolution equations, so the agreement 
of the calculations with experiment only depends essentially on the theoretical formulas fitting 
experiment at {\it one} value $Q_1^2$, say $Q_1^2=12\;\gev^2$, and on the validity 
of QCD for the subsequent evolution for larger $Q^2$ at small $x$.
 However, neither the hard nor the 
soft Pomeron solution are fully satisfactory. The hard Pomeron expression fails to fit data 
with $x\gsim0.01$. The soft Pomeron does not produce a marvellous fit either, 
and in addition presents conceptual problems, that we now briefly discuss.

First of all, we have the problem that, in the soft Pomeron case, the NLO 
overwhelms the LO term for small $x$ (as $\sim\alpha_s\sqrt{\log x}$), so the 
soft Pomeron-inspired formulas must necessarily 
fail for the strict $x\rightarrow 0$ limit. This is not 
the case for the hard singularity 
behaviour. Secondly, a power behaviour seems to be 
indicated for consistency with the $\gamma^*\gamma$ scattering 
case, where it has been shown to occur$^{[22]}$. Finally, we have the following argument. 
 By Reggeon calculus methods or summing ladder graphs, the 
authors of refs. 19, 20 have, at {\it tree} level, found the behaviour
$$F_S\sim x^{-\omega_0\alpha_s(\mu^2)},\;\omega_0=\dfrac{4C_A\log 2}{\pi}.$$
This poses no threat to the hard singularity behaviour since this term is 
subdominant with respect to $x^{-\lambda}$ if, as seems 
natural, the argument of $\alpha_s$ is proportional to $Q^2$; but it 
is incompatible with the soft Pomeron hypothesis because the new term dominates it. (However, 
there is a way out if the argument of $\alpha_s$ was the {\it hadronic} 
energy, $s\sim Q^2/x$ for then, as $x\rightarrow 0$, the new term  
would become merely a constant). 

A possibility that allows us to keep the best 
of both worlds is that one has, at a low $Q_0^2$,  a behaviour 
sum of the hard and soft Pomeron ones,
$$F_S(Q_0^2,x)\simeq c+b x^{-\lambda}.$$
Although this implies that in the limit $x\rightarrow 0$ the 
hard singularity will dominate, for finite $x$, if $c\ll b$, both 
soft and hard singularities may contribute comparable amounts. In this context 
it may be remarked that a behaviour like the one above has been shown$^{[23,1]}$ to 
describe very well photoproduction ($Q^2=0$) with constants precisely 
in the relation $c\ll b$. If this persists up to $Q_0^2\sim {\rm a\;few}\;\gev^2$, 
 the mixed behaviour would be indicated. Needless to say, since 
both soft and hard-singularity dominated 
behaviours fit the data a mixed one will do so even better: for 
example, the deficiencies of the hard singularity 
picture at ``large" $x$, and of the soft Pomeron one at all $x$, 
discussed in connection with the fit to the Zeus data, 
would likely be at least partially cured.\fonote{The hard singularity picture overshoots the 
large $x,\,Q^2$ data (Fig. 3) while the
 soft Pomeron one undershoots the small $x$ points, and undershoots large $x$ ones: see Fig. 6.} 
The conceptual difficulties of the soft Pomeron term alone also disappear as, 
in the strict limit $x\rightarrow 0$, it is dominated by the hard piece.

There exists also a theoretical argument in favour of the hard plus 
soft Pomeron situation, and it comes from  multi-Pomeron exchange theory. If, at 
a fixed $Q^2$ of the order of the GeV a single hard Pomeron gives\fonote{The following discussion is rather 
sketchy; details and references may be found in the review of ref. 24.}
$$\eqalign{F_{1P}(x,Q^2)\simeqsub_{s\rightarrow\infty}b_{1P}(Q^2)s^{\lambda},\cr
s\equiv Q^2/x,}$$
then an $n$-Pomeron term will produce the behaviour 
$$F_{nP}(x,Q^2)\simeqsub_{s\rightarrow\infty}b_{nP}(Q^2)s^{n\lambda};$$
the constants $b_{nP}$ should depend on the momentum at 
which they are calculated.

In some approximations (e.g., of eikonal type\ref{24}) one has, for $Q^2=-M^2_{\rm had}$, 
i.e., for on-shell scattering of hadrons,
$$b_{nP}(-M^2_{\rm had})=(-1)^{n+1}\dfrac{\kappa^n}{n!}C,$$
so we get for the sum
$$\eqalign{F_S(x,-M^2_{\rm had})=\sum_{n=1}^{\infty}F_{nP}(x,Q^2=-M^2_{\rm had})\cr
=C-C\exp\left[-\kappa s^{\lambda}\right]\simeqsub_{s\rightarrow\infty}C.}$$
For values of $Q^2$ of the order of $Q^2_0\sim 2\,-\,4\,\gev^2$, we expect that 
the $b_{nP}(Q^2)$ will not change much from $b_{nP}(-M^2_{\rm had})$, so if we write
$$b_{nP}(Q_0^2)\simeq b_{nP}(-M^2_{\rm had})+\Deltav_n,$$
we will then get,
$$F_S(x,Q_0^2)\simeq F_S(x,-M^2_{\rm had})+\sum \Deltav_n\simeq C+\Deltav_1 x^{-\lambda}+
\Deltav_2 x^{-2\lambda}+\dots ,\eqno (5.1)$$
with the $\Deltav$ small. This expression contains a hard plus a soft Pomeron of 
the type discussed above. It also contains a
 term $\Deltav_2 x^{-2\lambda}$, 
whose inclusion we will consider in \sect 5.3.  
 \vskip.1cm
{\bf 5.2. Hard plus soft singularities for $F_2$.}
\vskip.2cm
 We next discuss the large $Q^2\gsim 10\,\gev^2$ 
region, under various hypotheses for the small $Q^2$ region, which we then evolve with QCD. 
We will consider moderately large values of $x$,  
 $x\leq 0.032$ because we will be interested not 
only on the leading behaviour as $x\rightarrow 0$, given almost certainly 
by a hard singularity, but also on the subleading corrections.
 So we assume that at a certain, fixed $Q^2_0\sim 1\,\gev^2$, one has
$$F_S(x,Q^2_0)\simeq B_Sx^{-\lambda}+F_{\rm corr.}(x,Q^2_0).\eqno (5.2)$$
For the correction term, $F_{\rm corr.}(x,Q^2_0)$, we consider the following 
possibilities: a soft Pomeron,
$$F^P_{\rm corr.}(x,Q^2_0)\simeq {\rm constant};\eqno (5.3{\rm a})$$
and a $P'$ Regge pole,\fonote{For Regge pole theory cf. ref. 25. }
$$F^{P'}_{\rm corr.}(x,Q^2_0)\simeq {\rm constant}\times x^{1-\alpha_{P'}(0)},
\alpha_{P'}(0)\sim 0.5 .\eqno (5.3{\rm b})$$
This last possibility is considered because, as shown 
in ref. 10, any behaviour $x^\sigma,\,\sigma\geq 0$ at $Q_0^2$, produces at 
larger $Q^2$ behaviours differing from the soft Pomeron one, (1.4a), {\it only 
in the pre-factor}, but with the same exponent. 

Once assumed the behaviours given in (5.3), and taking 
for simplicity that the gluon
 structure function behaves like the quark singlet one, we 
evolve with QCD for higher $Q^2$. From the results of the previous \sects  
we find, to NLO,
$$F_S(x,Q^2)\simeq 
B_S\left\{1+\dfrac{c_S(1+\lambda_0)\alpha_s}{4\pi}\right\}
\ee^{q_S(1+\lambda_0)\alpha_s/4\pi}
 [\alpha_s(Q^2)]^{-d_+(1+\lambda_0)}x^{-\lambda_0}+F_{\rm corr.}(x,Q^2),\eqno (5.4)$$
 and 
depending on the low $Q^2_0$ hypothesis we make we find the following correction terms:
$$\eqalign{F^P_{\rm corr.}(x,Q^2)\simeq \left\{1+2\sqrt{\dfrac{\xi}{ d_0\tau}}\left[k_1\;
\dfrac{\alpha_s(Q^2)}{4\pi}-
k\;\dfrac{\alpha_s(Q^2_0)}{4\pi}\right]\right\}\cr
\times \frac{c_0}{\xi}\left[ \frac{9\xi \tau}{4\pi^2(33-2n_f)}
\right]^{\frac{1}{4}}
\exp\left\{ \sqrt{d_0\xi\;
\tau}-
d_1\tau \right\}}\eqno (5.5{\rm a})$$
for a soft Pomeron, \eq (5.3a). 
 For the $P'$ Regge pole we find a very 
similar formula:
$$\eqalign{F^{P'}_{\rm corr.}(x,Q^2)\simeq 
\left\{1+2\sqrt{\dfrac{\xi}{ d_0\tau}}\left[k_1\;
\dfrac{\alpha_s(Q^2)}{4\pi}-
k\;\dfrac{\alpha_s(Q^2_0)}{4\pi}\right]\right\}\cr
\times\dfrac{ c_{P'}}{\sqrt{\xi}}\left(\dfrac{\tau}{\xi}\right)^{\frac{3}{4}}\exp\left\{\sqrt{d_0\xi\;
\tau}-
d_1\tau \right\}.}
\eqno (5.5{\rm b})$$

As we will see, none of the three possibilities gives a really good fit 
in the ``large" $0.01<x\leq 0.032$ region; for the more 
precise Zeus data the \chidof  is of 1.7. To remedy this we consider the 
possibility of softening the large $x$ region by multiplying $F_S(x,Q^2)$ by a factor 
$(1-x)^{\nu}$, as discussed in \sect 3. Here, however, we take $\nu$ constant because 
this already produces an excellent fit.

The results are summarized in Table IX, where for definiteness 
we compare the fits obtained with (5.5a,b) with 
the fits found  using only the soft Pomeron-dominated 
expression. We have {\it not} fitted $\lambda$, which we have set equal to 0.470, for 
reasons that will be apparent in next section; if we had fitted it, we would have obtained 
$\lambda=0.43$ and an improvement of only two units in the chi-squared\fonote{To be precise,
 if we had {\it fitted} $\lambda$ 
to e.g., the Zeus data using a hard plus soft term, we would 
have obtained (not correcting for large $x$, and fixing $\Lambdav=0.23\,\gev$),
$$\lambda=0.429,\;Q_0^2=2.40\;\gev^2,\; \langle e^2_q\rangle c_0=0.244,
\; \langle e^2_q\rangle B_S=3.83\times10^{-4},\; \langle e^2_q\rangle B_{NS}=0.30,$$
for a \chidof $=\tfrac{195}{120-5}$. 
}.
We give the results for the Zeus data only; later we 
will present simultaneous fits to H1 and Zeus data.
\vskip.2cm
\setbox0=\vbox{\hsize=15.cm
{\petit 
\vskip.2cm
\centerline{{\bf Table IX}.- $n_f=4$; Zeus data, plus neutrino data. $x\leq 0.025,\,Q^2\geq 12.5\,\gev^2$}
\centerrule{16em}
\vskip.1cm
 $$\matrix{
{\rm Soft\;Pomeron\;only}\left\{
\matrix{\Lambdav & Q_0^2& \langle e^2_q\rangle c_0 & \langle e^2_q\rangle  B_{NS}&\chi^2/{\rm d.o.f.}\cr
0.165\;\gev&0.90\,\gev^2&0.282&0.240&\tfrac{273}{120-4}}\right.\cr
&\phantom{x} & & & &\cr
{\rm Hard+Soft\;Pomeron^*}\left\{
\matrix{\lambda\;({\rm fixed})& Q_0^2& \langle e^2_q\rangle c_0 &
\langle e^2_q\rangle B_S& \langle e^2_q\rangle  B_{NS}&\chi^2/{\rm d.o.f.}\cr
0.47&2.45\;\gev^2&0.252&4.42\times10^{-4}&0.294&\tfrac{197}{120-4}}\right.
\cr
&\phantom{x} & & & &\cr
\matrix{{\rm Hard}+P'^*,\cr
 {\rm large\;}x\;{\rm softened}^{**}} \left\{
\matrix{\lambda\;({\rm fixed})& Q_0^2& \langle e^2_q\rangle c_{P'} &
\langle e^2_q\rangle B_S& \langle e^2_q\rangle  B_{NS}&\chi^2/{\rm d.o.f.}\cr
0.47&1.11\;\gev^2&0.616&4.25\times10^{-4}&0.343&\tfrac{171}{120-4}}\right.\cr
&\phantom{x} & & & &\cr
\matrix{{\rm Hard+Soft^*,}\cr 
{\rm large\;}x\;{\rm softened}}\left\{
\matrix{\lambda\;({\rm fixed})& Q_0^2& \langle e^2_q\rangle c_0 &
\langle e^2_q\rangle B_S& \langle e^2_q\rangle  B_{NS}&\chi^2/{\rm d.o.f.}\cr
0.47&2.72\;\gev^2&0.310&3.417\times10^{-4}&0.370&\tfrac{143}{120-4}}\right.}
$$
\vskip.1cm
\noindent $^*\;\Lambdav\;{\rm fixed\;at}\;0.230\;\gev.$
 The optimum value would correspond to $\Lambdav\sim 0.45\;\gev$.\hb
$^{**}$ If we had not corrected for the large $x$ values, i.e., we had not included the 
factor $(1-x)^\nu$, we would have obtained a \chidof  of 270.}
}
\centerline{\boxit{\box0}}            
\vskip.2cm
In this table the expression ``large $x$ softened" means that we have multiplied the 
formulas for $F_S$ by a factor $(1-x)^\nu,\,\nu\simeq 11$, to correct the structure functions for (relatively) 
large values of $x$.  
 For the hard singularity case, cf. ref. 5 and \sect 3.4 here; for the Hard + Soft singularities 
case, we have taken $\nu=10\sim 11$. 
(We will discuss further the ``large" $x$ region in \sect 5.3).
 \vskip.1cm
{\bf 5.3. Hard plus soft singularities, plus triple Pomeron term for $F_2$. Best (global) fits.}
\vskip.2cm
 
Clearly, the best fit is obtained with the hard plus soft Pomerons. Not only the \chidof 
is quite good, but the values of the parameters are 
very reasonable. In fact, more evidence in favour of the ``hard plus soft" scenario  
will be given in next section; for now we will consider that the fits are so good, 
that it makes sense to see if one can find evidence for a ``triple Pomeron" term. That is, we 
consider that [cf.  Eq. (5.1)] 
$$F_S=F_{\rm Soft}+F_{\rm Hard}+F^{TP},$$
$$F^{TP}(x,Q^2_0)\simeq ({\rm Const.})\,x^{-2\lambda},$$
so that, when evolved with QCD to large $Q^2$,
$$F^{TP}(x,Q^2)\simeq  
B_{TP}\left\{1+\dfrac{c_S(1+2\lambda)\alpha_s}{4\pi}\right\}
\ee^{q_S(1+2\lambda)\alpha_s/4\pi}
 [\alpha_s(Q^2)]^{-d_+(1+2\lambda)}x^{-2\lambda}.\eqno (5.6)$$
Note that this is $O(\alpha_s^{d_+(1+\lambda)-d_+(1+2\lambda)})\simeq O(\alpha_s^{1.4})$, 
 i.e., subleading in powers of $\alpha_s$, with respect to $F_{\rm Hard}$.

We present in Table X the parameters of 
the fits to, simultaneously, H1 and Zeus data on $ep$, plus neutrino data. This gives 
our best set of formulas, providing an excellent
 fit to experiment in a very wide range of $Q^2,\,x$.
 In the second case (Table X$b$) we do not give the fit including a triple Pomeron term 
as the \chidof  does not vary appreciably if including it provided 
$ \langle e^2_q\rangle |B_{TP}|\lsim 2\times 10^{-4}$. We consider
 the parameters given in Table X$a$ and Table X$c$ (see below)  
to be the more reliable ones for describing small $x$ structure functions. If we had 
fitted also $\lambda$ with the whole set of data we would have obtained minima 
for values comprised between 0.42 and 0.49, with a variation of 
the chi-squared of less than two units with respect to the one obtained 
fixing $\lambda=0.470$. Finally, if we fit the QCD 
parameter $\Lambdav$, the values 
 which provide minima vary between 0.555 \gev$\,$  and 0.310 \gev, and the chi-squared 
improves by less than five units. Because of this we 
consider, as stated, that it is justified to favour the 
fits obtained with {\it fixed} $\lambda=0.470,\,\Lambdav=0.23\,\gev$.
\vskip.2cm
\setbox0=\vbox{\hsize=14.cm
{\petit 
\vskip.2cm
\centerline{{\bf Table X$a$}. $n_f=4$; Zeus plus H1 data; $Q^2\geq 10\,\gev^2,\,x\leq 0.01$.}
\centerrule{16em}
\vskip.1cm
 $$\matrix{{\rm Hard}+P \left\{
\matrix{\lambda\;({\rm fixed})& Q_0^2& \langle e^2_q\rangle c_{P} &
\langle e^2_q\rangle B_S& \langle e^2_q\rangle  B_{NS}&
\chi^2/{\rm d.o.f.}\cr
0.47&2.95\,\gev^2&0.296&4.28\times10^{-4}&0.349&\tfrac{138.3}{144-4}}\right.\cr
\matrix{{\rm Hard}+P,\cr
 +{\rm TP\; term}} \left\{\matrix{\lambda\;({\rm fixed})& Q_0^2& \langle e^2_q\rangle c_{P} &
\langle e^2_q\rangle B_S& \langle e^2_q\rangle B_{TP}& \langle e^2_q\rangle  B_{NS}&
\chi^2/{\rm d.o.f.}\cr
0.47&4.45\,\gev^2&0.258&8.33\times10^{-4}&-1.67\times10^{-4}&0.359&\tfrac{129.3}{144-5}}\right.}
$$
}            
\vskip.2cm
\hrule
\vskip.2cm
{\petit 
\vskip.2cm
\centerline{{\bf Table X$b$}. $n_f=4$; Zeus plus H1 data;
 $Q^2\geq 10\,\gev^2,\,x\leq 0.032$.}
\centerrule{16em}
\vskip.1cm
 $$\matrix{{\rm Hard}+P\cr
 $x$\;{\rm ``softened"\; with}\; (1-x)^\nu}
 \left\{
\matrix{\lambda\;({\rm fixed})& Q_0^2& \langle e^2_q\rangle c_{P} &
\langle e^2_q\rangle B_S& \langle e^2_q\rangle  B_{NS}&
\chi^2/{\rm d.o.f.}\cr
0.47&2.28\,\gev^2&0.311&2.72\times10^{-4}&0.315&\tfrac{227.4}{230-4}}\right.
$$
}
\vskip.2cm
\hrule
\vskip.2cm
{\petit 
\vskip.2cm
\centerline{{\bf Table X$c$}. $n_f=4$; Zeus plus H1 data;
 $Q^2\geq 10\,\gev^2,\,x\leq 0.032$..}
\centerrule{16em}
\vskip.1cm
 $$\matrix{{\rm Hard}+P\cr
+P'} \left\{
\matrix{\lambda\;({\rm fixed})& Q_0^2& \langle e^2_q\rangle c_{P}& \langle e^2_q\rangle C_{P'} &
\langle e^2_q\rangle B_S& \langle e^2_q\rangle  B_{NS}&
\chi^2/{\rm d.o.f.}\cr
0.47&5.00\,\gev^2&0.588&-0.271 &4.66\times10^{-4}&0.262&\tfrac{265.3}{230-5}}\right.
$$
\vskip.1cm
\hrule
\vskip.1cm
\centerline{$\Lambdav\;{\rm fixed\;at}\;0.230\;\gev.$ NLO corrections included}}
}
\centerline{\boxit{\box0}}            
\vskip.2cm

We discuss now in some detail the larger $x$ region. If taken by themselves, both soft 
and hard Pomeron expressions (and {\it a fortiori} a sum of the two) 
must, as discussed in ref. 1 and \sect 3.1 here, 
 run in contradiction with the momentum sum rule if mantained 
for fixed $x$ and $Q^2\rightarrow \infty$; and this contradiction starts becoming noticeable at 
the higher $Q^2\sim1\,000\,\gev^2$ for $x\geq 0.02$: so 
a modification of our formulas for finite $x$ is necessary. In the present paper  
we have, until now, introduced it phenomenologically by multiplying the low $x$ expressions by 
a factor $(1-x)^\nu$ (``softening"), with $\nu$ constant or depending on $Q^2$. A more
 rigorous procedure would be to assume, at a {\it fixed} $Q^2_0$, 
a behaviour like
$$F_S(x,Q^2_0)= (\bar{B}_Sx^{-\lambda}+C)(1-x)^{\nu_0},\eqno (5.7)$$
and then evolve with QCD. This is best done by expanding first (5.7) in $x$,
$$F_S(x,Q^2_0)\simeq \bar{B}_Sx^{-\lambda}+C-\nu_0B_Sx^{1-\lambda}+\dots.\eqno (5.8)$$
The dots correspond to terms behaving, as $x\rightarrow 0$, as higher powers 
of $x$, which need not be considered; see below. (5.8) has
 exactly the form of a hard Pomeron, plus a soft 
Pomeron, plus a $P'$ Regge pole, plus higher powers of $x$.
 From the results of the previous sections we know that 
all terms vanishing for small $x$ yield the same expression, {\it up to a constant}, when evolved 
to large $Q^2$, as the $P'$ piece. So we may lump the piece $-\nu_0B_Sx^{1-\lambda}+\dots$ 
into a single term like that of \eq (5.5b), to be added to a soft and a hard 
term. To LO thus,
$$F_S(x,Q^2)\simeqsub_{{x\rightarrow 0}\atop{Q^2\rightarrow\infty}}
B_S\alpha_s^{-d_+}+F^P_{\rm corr.}(x,Q^2)+F^{P'}_{\rm corr.}(x,Q^2)\eqno (5.9)$$
$F^P_{\rm corr.}(x,Q^2),\,F^{P'}_{\rm corr.}(x,Q^2)$ given in \eqs (5.5). 
 The resulting fit (including NLO corrections) is described in Table X{\it c} above.
 This certainly improves the fit at large $x$ with respect 
to the unsoftened situation, but not much; indeed, less than the simple 
``softening" used before in the text\fonote{Nevertheless, from the point of view of {\it rigorous} QCD the softening given 
in \eq (5.9) is to be preferred to mere multiplication by $(1-x)^\nu$ for all $Q^2$; e.g., for 
extrapolations to higher $Q^2$, since (5.9)  
is compatible with QCD evolution for $x\rightarrow 0$. What is more, the parameters are 
fairly stable from Table X$a$ ot Table X$c$, advantages that in our opinion offset 
a small increase in \chidof. wit respect to Table X$b$.}, and it certainly does not solve 
the momentum sum rule problem, either. This should not be too surprising: by 
its very nature, a calculation with leading terms only 
for $x$ must fail for larger values of this variable.

What 
we mean by this exactly is the following. Consider that at a 
fixed $Q_0^2$ one had exactly a hard Pomeron, $F_S(x,Q_0^2)=\bar{B}x^{-\lambda}$. Then, at any 
larger $Q^2$, we have the moments [in this simplified discussion 
we neglect the matrix character of the evolution equations]
 $$\mu(n,Q^2)=\bar{B}\left[\dfrac{\alpha_s(Q^2_0)}{\alpha_s(Q^2)}\right]^{d_+(n)}
\dfrac{1}{n-(1+\lambda)}.$$
 Writing identically
 $\ee^{\tau d_+(n)}\equiv \ee^{\tau d_+(1+\lambda)}[1+\delta(n,\lambda)]$ 
we find
$$F_S(x,Q^2)=F_S^{\rm Lead.}+F_S^{\rm SL};\;
F_S^{\rm Lead.}(x,Q^2)=B_S[\alpha_s(Q^2)]^{-d_+(1+\lambda)}x^{-\lambda},$$
and the subleading piece is such that it has moments
$$\mu^{\rm SL}(n,Q^2)=B_S[\alpha_s(Q^2)]^{-d_+(1+\lambda)}
\dfrac{\delta(n,\lambda)}{n-(1+\lambda)}.$$
$\delta(n,\lambda)$ vanishes for $n=1+\lambda$; hence the first 
singularity of the $\mu^{\rm SL}(n,Q^2)$ occurs for $n=1$

 and there,
$$\mu^{\rm SL}(n,Q^2)\simeqsub_{n\rightarrow 1}
\dfrac{-B_S}{1+\lambda}\alpha_s^{-d_+(1+\lambda)}\exp\tau\left[\dfrac{d_0}{4(n-1)}-d_1\right].$$
This is of the soft-Pomeron type apart from the factor $\alpha_s^{-d_+(1+\lambda)}$,
 so we expect
$$F_S^{\rm SL}(x,Q^2)\sim\alpha_s^{-d_+(1+\lambda)+d_1}\ee^{\sqrt{d_0\tau\xi}}.$$
If we continie to subtract the tems singular at $n=1,\,2,\dots$, we would get an 
asymptotic series for $F_S$. 
For $x\rightarrow 0$ this is dominated 
by the term $F_S^{\rm Lead.}$; but for fixed $x$, the remaining terms in the series
 end up by overwhelming 
 $F_S^{\rm Lead.}$ as $Q^2$ becomes {\it very} large.

A precise evaluation is not difficult; it would also involve the gluon component. We shall not 
present the corresponding fits here; to do so one would have to include 
also subdominant corrections to the {\it soft} Pomeron piece. It is unclear that the 
effort would be worth the results, given the good quality
 of the fits with dominant terms only. 

\setbox0=\vbox{\hsize 13.5truecm \epsfxsize=13.4truecm\epsfbox{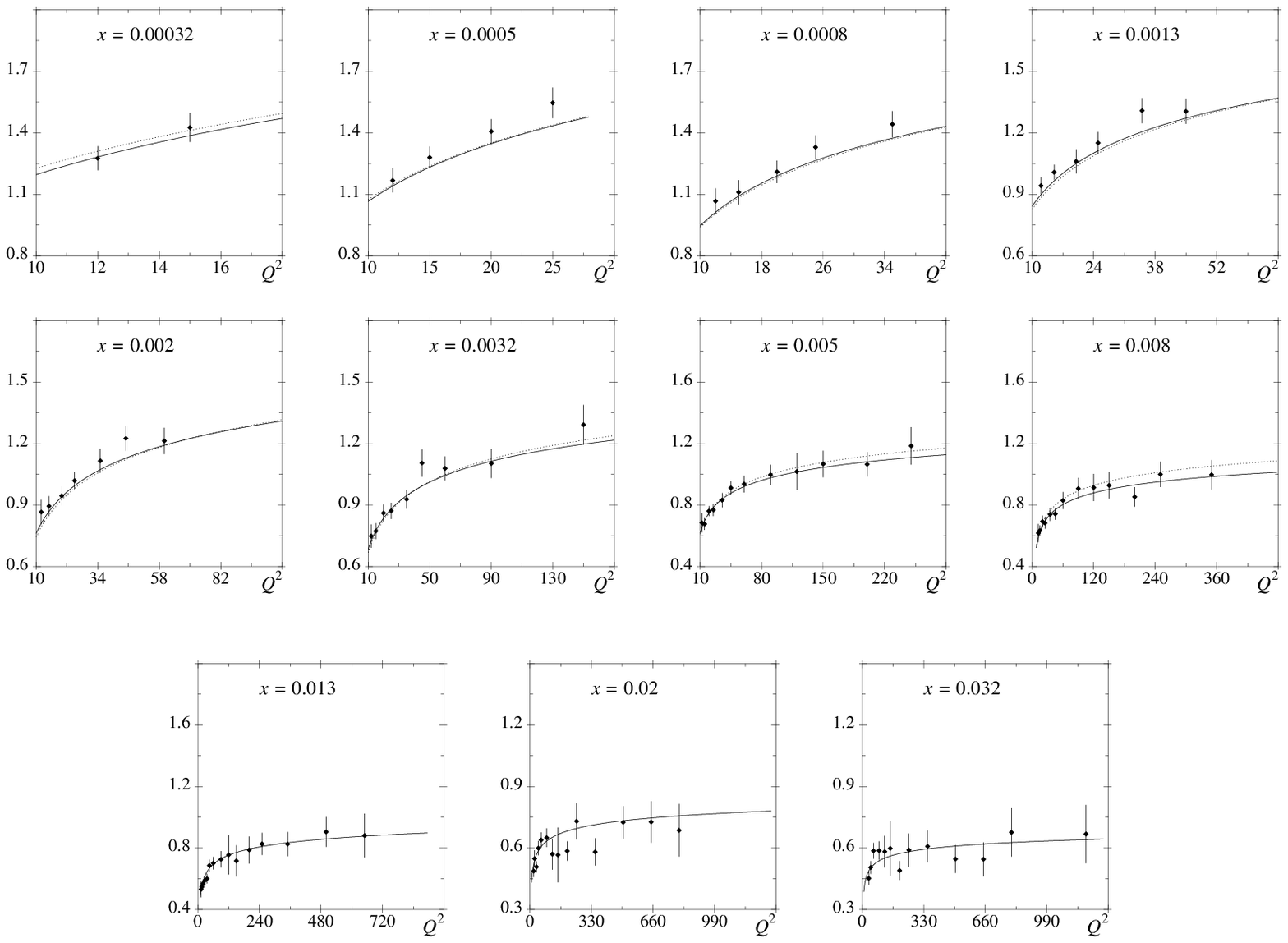}\hb{\petit
\noindent Figure 8a. Comparison of predictions from 
 Table X with H1 $ep$ data for $F_2$.
\vskip.1cm}}
\centerline{\box0}
\vskip.1cm
\setbox0=\vbox{\hsize 13.5truecm \epsfxsize=13.4truecm\epsfbox{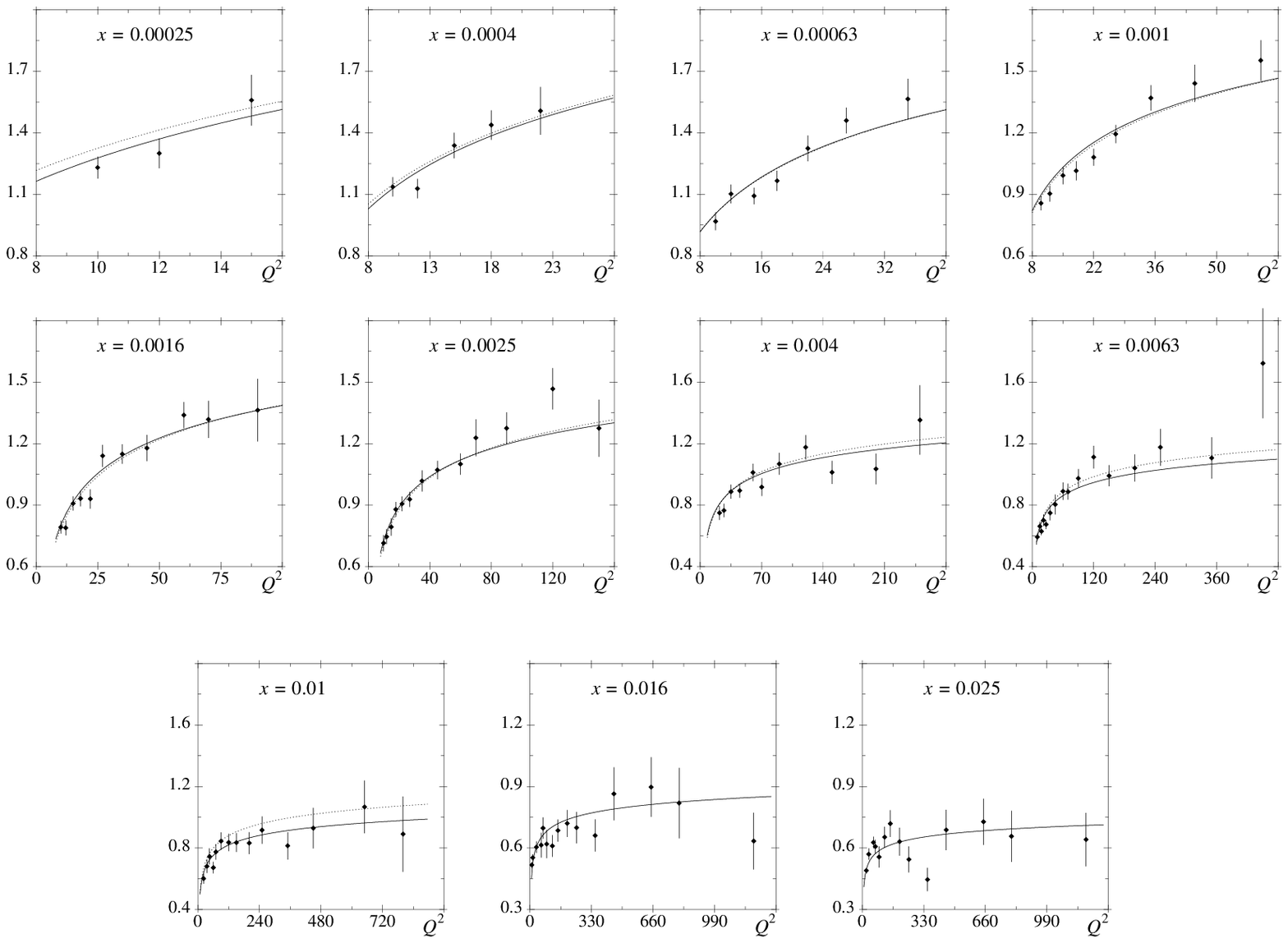}\hb{\petit
\noindent Figure 8b. Comparison of predictions from 
 Table X with Zeus $ep$ data for $F_2$.
\vskip.2cm}}
\centerline{\box0}

In Figs. 8 we show the comparison of our fits, with the parameters in Table X, with data. Note 
that the {\it same} values of the parameters are used in  Fig. 8a  and Fig. 8b.
 For both Figs. 8 we give the fits with the 
``softened" and straight formulas:
 the  continuous lines indicate large-$x$ softening, and the dotted 
lines no softening. The improvement in the quality of the fits when
 compared with the ``soft-Pomeron only" or ``hard-Pomeron only"
 is obvious to the naked eye here.
\vskip.1cm
{\bf 5.3. Gluon and longitudinal structure functions.}
\vskip.2cm
Detailed predictions
 for the gluon and longitudinal structure functions are obtained trivially 
by adding the soft and hard Pomeron expressions given in the previous sections  We 
will leave the details of this to the reader; likewise, we do not draw the 
figures for the $F_G,\,F_L$ as they would not differ much from the ones 
drawn for, say, the hard Pomeron-only hypothesis. However 
we would like to comment here on a particularly interesting prediction of
 our analysis for the growth of  
 the cross-section $\sigma_{\gamma p\rightarrow J/\psi p}(W)$ as 
a function of the c.m. energy, $W$. In fact, this cross section may be expressed 
as a function of the gluon structure function $F_G$, 
$$\sigma_{\gamma p\rightarrow J/\psi p}(W)=AF_G(\bar{x}=
a\dfrac{M_{J/\psi}^2}{W^2}, Q^2=M_{J/\psi}^2)\simeq 9\,\gev^2,$$
and $A,\,a$ are constants approximately known. For $F_G$ we have, 
$$F_G(x,Q^2)\simeqsub_{x\rightarrow 0} B_G[\alpha_s]^{-d_+}x^{-\lambda}$$
and $B_G$ may be calculated in terms of $\lambda,\,B_S$. So, using our 
formulas we have, for the logarithmic 
{\it slope} of the cross section,
$$\delta\equiv\dfrac{\log\sigma_{\gamma p\rightarrow J/\psi p}(W)}{\log W}
\rightarrowsub_{s\rightarrow \infty}
 2\lambda.$$
The figure reported in a
 fit\ref{26} including recent HERA data\ref{27}  gives $\delta=2\lambda=0.9$. This 
in very good agreement with the optimum values of $\lambda$ obtained 
with the hard plus soft fits, $2\lambda=0.83\;{\rm to}\,1.0$, and is clearly superior 
to the results following from the hard Pomeron only hypothesis, 
$2\lambda=0.64\;{\rm to}\,0.76$.
\vskip.4cm
\centerline{6. {\bf HARD PLUS SOFT SINGULARITIES: SMALL $Q^2$.}}
\vskip0.5cm
The quality of the results obtained by assuming that at values of 
$Q_0^2\sim 3\,\gev^2$ one has a hard singularity, $x^{-\lambda}$, plus a soft (constant) 
Pomeron term, evolved with QCD to large values $Q^2\geq 10\,\gev^2$, leads us 
naturally to the question wether it is possible to extend the 
analysis to the {\it low} $Q^2$ region as well, thus   
enabling us to address the important issue of the connection between 
 the perturbative regime ($Q^2\geq 10\,\gev^2$, say) and the region
 $Q^2<10\,\gev^2$ 
where nonperturbative effects are determinant.

It should be obvious that, unless one were able to perform a full, nonperturbative 
calculation, we must content ourselves with {\it phenomenological} evaluations. Here 
we use approximate, QCD-inspired formulas and assumptions
 and enquire wether we can still fit the data. We will find that this is indeed the case; in 
particular, we will see that the extension of the fit of the data to $Q^2\rightarrow 0$ implies 
self-consistency conditions both for the singlet and 
the nonsinglet which will allow us to {\it calculate} the constants $\lambda,\,\rho$, 
getting values in impressive agreement with other (in particular, high $Q^2$) determinations.

 The expression for the virtual photon scattering
 cross section in terms of 
the structure function $F_2$ is
$$\sigma_{\gamma(Q^2=0)p}(s)=\dfrac{4\pi\alpha}{Q^2}F_2(x,Q^2),\;{\rm with}\; s=Q^2/x.
\eqno (6.1)$$
We would like to describe this down to $Q^2\rightarrow 0$. 
In the low energy region we should, as discussed, take the soft-Pomeron 
dominated expression to be given by an ordinary Pomeron, i.e. ,
 behaving as a constant for $x\rightarrow 0$ (or equivalently, $s\rightarrow \infty$): 
 the expression for $F_2$ that will, when evolved to large $Q^2$ 
yield (5.4), (5.5a) is
$$\eqalign{F_2=\langle e_q^2\rangle\Big\{
B_S[\alpha_s(Q^2)]^{-d_+(1+\lambda_0)}x^{-\lambda_0}\cr
+C+B_{NS}[\alpha_s(Q^2)]^{-d_{NS}(1-\rho_0)}x^{\rho_0}\Big\}.}
\eqno (6.2)$$
Because NLO corrections are large for $Q^2\leq 10\,\gev^2$ and 
we are interested in a semi-phenomenological description, only LO formulas will 
be used. 
Note also that the $C$ in \eq (6.2) is different from the $c_0$ in, 
say, (1.4a) as the gluon component also intervenes
 in the evolution.

On comparing (6.1) and (6.2) we see that, as noted in ref. 1, we have 
problems if we want to extend (6.2) to very small $Q^2$. First of all, 
$$\alpha_s(Q^2)=\dfrac{4\pi}{\beta_0\log Q^2/\Lambdav^2}
\eqno (6.3)$$ {\it diverges} when $Q^2\sim\Lambdav^2$. Secondly, 
Eq. (6.1) contains the factor $Q^2$ in the denominator so the 
cross section blows up as $Q^2\rightarrow 0$ unless $F_2$ were to develop 
a zero there.

It turns out that there is a simple way to solve both 
difficulties at the same time. It has been conjectured$^{[28]}$ that the 
expression (6.3) for $\alpha_s$ should be modified for values of 
$Q^2$ near $\Lambdav^2$ in such a way that it {\it saturates}, producing in particular
 a finite value for $Q^2\sim\Lambdav^2$. To be precise, one alters (6.3) according to
$$\alpha_s(Q^2)\rightarrow\dfrac{4\pi}{\beta_0\log (Q^2+M^2)/\Lambdav^2},$$
where $M$ is a typical hadronic mass, $M\sim m_\rho\sim\Lambdav(n_f=2)\dots$; 
 the value $M=0.96\,\gev$ has been suggested on the basis 
of lattice calculations. It has been argued 
that saturation incorporates important nonperturbative effects.
 In the present paper we will simply set $M=\Lambdav=\Lambdav_{\rm eff}$, to avoid 
a proliferation of parameters. For the Pomeron term [the constant in Eq. (6.2)] 
we merely replace $C\rightarrow Q^2/(Q^2+\Lambdav^2_{\rm eff})$, using 
a procedure similar to that of ref. 29. The  
expression we will use for low $Q^2$ is thus,
$$\eqalign{F_2=\langle e_q^2\rangle\Big\{
B_S[\tilde{\alpha}_s(Q^2)]^{-d_+(1+\lambda)}Q^{-2\lambda}s^\lambda\cr
+C\dfrac{Q^2}{Q^2+\Lambdav^2_{\rm eff}}+
B_{NS}[\tilde{\alpha}_s(Q^2)]^{-d_{NS}(1-\rho)}Q^{2\rho}s^{-\rho}\Big\},}
\eqno (6.4{\rm a})$$
where
$$\tilde{\alpha}_s(Q^2)=\dfrac{4\pi}{\beta_0\log (Q^2+\Lambdav_{\rm eff}^2)/\Lambdav_{\rm eff}^2}
\eqno (6.4{\rm b})$$
and we have changed variables, $(Q^2,\,x)\rightarrow (Q^2,\,s=Q^2/x)$.

We have still not solved our problems: given Eq. (6.1) it is clear that 
a {\it finite} cross section for $Q^2\rightarrow 0$ will 
only be obtained if the powers of $Q^2$ match exactly. This is automatic by 
construction for the Pomeron term, but for the hard singlet and the nonsinglet 
piece it will only occur if we have consistency conditions satisfied. With 
the expression given in (6.4b) for $\tilde{\alpha}_s$ it diverges 
as ${\rm Const.}/Q^2$ when $Q^2\rightarrow 0$: so we only get 
a matching of zeros and divergences for $\sigma_{\gamma(Q^2=0)p}(s)$
 if $\lambda=\lambda_0,\,\rho=\rho_0$ such that   
$$d_+(1+\lambda_0)=1+\lambda_0,\;d_{NS}(1-\rho_0)=1-\rho_0.\eqno (6.5)$$
The solution to these expressions depends very little on the number of
 flavours; for $n_f=2$, probably 
the best choice at the values of $Q^2$ we will be working with, one finds 
$\lambda_0=0.470,\,\rho_0=0.522$. The second is in uncanny 
agreement with the value obtained with either a Regge 
analysis in hadron scattering processes, or 
by fitting structure functions in DIS. The first is 
 larger than the value obtained in the fits to DIS with {\it only} a hard Pomeron, which gave  
$\lambda=0.32\;{\rm to}\;0.38$; but falls in the right ballpark of values 
obtained in the previous section with hard plus 
soft Pomeron, $\lambda=0.43\,{\rm to}\,0.5$.

\setbox3=\vbox{\hsize 15.1truecm
\setbox1=\vbox{\hsize 15.truecm \epsfxsize=15.truecm\epsfbox{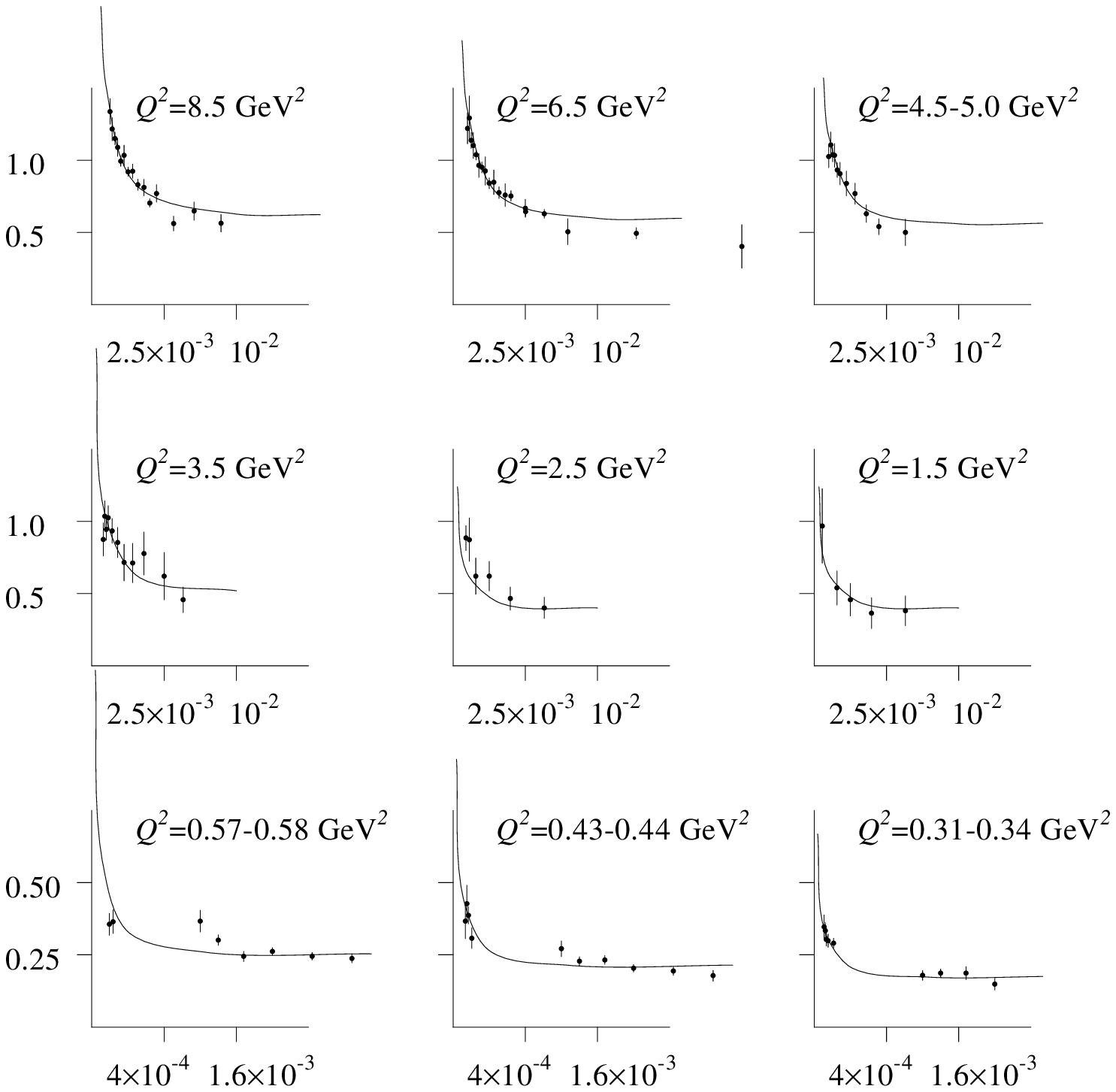}}
\centerline{\box1}
\vskip-.6cm
\setbox0=\vbox{\hsize12.1truecm \petit \noindent Figure 9. Comparison of predictions
 with data, Zeus plus H1. The
neutrino data and prediction are not shown. We plot $F_2$ {\it vs.} $x^{1/2}$.\hb
\vskip.2cm} 
\line{\box0}}
\centerline{\box3}

We are perfectly aware that, by using \eqs(6.4) down to 
$Q^2=0.32\,\gev^2$ we are pushing perturbative QCD well below its region 
of applicability, and that the condition of matching at $Q^2\rightarrow 0$ is 
at best only of phenomenological value. Nevertheless, the fact that we get 
such reasonable predictions for $\lambda_0,\,\rho_0$ probably 
indicates that our procedure represents, {\it grosso modo}, 
the actual situation, which is also justified by  
the quality of the 
fit Eq. (6.4) provides. If we take all H1 and Zeus data for
 $0.31\,\gev^2\leq Q^2\leq 8.5\,\gev^2$, and we include also  
10 neutrino $xF_3$ data we 
find
$$\Lambdav_{\rm eff}=0.87\, \gev,\;\langle e_q^2\rangle B_S= 5.28\times10^{-3},
\;\langle e_q^2\rangle B_{NS}=0.498,\;\langle e_q^2\rangle C=0.486,$$
for a \chidof$=\tfrac{106.2}{104-4}$.
The value of $\Lambdav_{\rm eff}$ we have obtained lies somewhere inside the expected 
bracket, $\Lambdav(n_f=2)\simeq 0.35\,\gev$ and the value found in the quoted lattice 
calculation for $M$, 0.96 GeV. Clearly, the fit gives a compromise, phenomenological quantity.

 The agreement 
between phenomenology 
and experiment, shown graphically in Fig. 9, is unlikely to be trivial; $x$ varies between
 $6\times 10^{-6}$ and $4\times 10^{-2}$ , and $F_2$ changes by almost one order 
of magnitude. To see more clearly this nontriviality, we replace the 
hard singularity by an evolved soft Pomeron, with a saturated $\alpha_s$. That is, 
we now fit with the expression

$$\eqalign{F_2=\langle e_q^2\rangle 
\Bigg\{\frac{c_0}{\xi}
\left[ \frac{9\xi \log[\tilde{\alpha}_s(Q_0^2)/\tilde{\alpha}_s (Q^2)]}{4\pi^2(33-2n_f)}
\right]^{\frac{1}{4}}
\exp\left( \sqrt{d_0\xi\;
\left[\log\frac{\tilde{\alpha}_s(Q_0^2)}{\tilde{\alpha}_s(Q^2)}\right]}-
d_1\log\frac{\tilde{\alpha}_s(Q_0^2)}{\tilde{\alpha}_s(Q^2)} \right)
\cr
+C\dfrac{Q^2}{Q^2+\Lambdav^2_{\rm eff}}+
B_{NS}[\tilde{\alpha}_s(Q^2)]^{-d_{NS}(1-\rho)}Q^{2\rho}s^{-\rho}\Bigg\},}$$
$\tilde{\alpha}_s$ as before. Then we find
$$\Lambdav_{\rm eff}=0.41\,\gev,\;\langle e_q^2\rangle c_0=0.094,\;\langle e_q^2\rangle C=0.253,
\;\langle e_q^2\rangle B_{NS}=0.41$$
and a much deteriorated \chidof $=250/(104-4)$. We consider these results as convincing proof of the 
necessity of a hard component also at low $Q^2$.
 
\vskip.4cm
\centerline{\bf 7. DISCUSSION}
\vskip.5cm 
The main outcome of our 
analysis in the present paper is that we are able to give a unified,
 consistent description of small $x$ DIS data, both for large and small values of $Q^2$, 
by assuming, at low momenta, the presence of a hard plus a soft Pomeron, a procedure which 
improves substantially the quality of the descriptions with only one of these. There also  
appears some evidence 
for a triple Pomeron contribution; evidence which is, however, somewhat marginal.  
Besides this, there are a number of specific points to which we would
 like to draw also attention.

 First of all there is the matter of the dependence of 
our low $Q^2$ results on the saturation hypothesis for $\alpha_s$. It is clear 
that the good quality of the fits indicates that, 
with suitable modifications, perturbative QCD supplemented by saturation, 
 may give a {\it phenomenological} 
description of the data down to very low momenta; but of course this should not be construed 
as a proof of saturation, in particular of the very specific form considered 
here. One may interpret our results, however, as showing that the saturation expression 
is particularly adapted to represent, in DIS, a variety of effects: higher 
twists, renormalons, and likely also genuine saturation.

A second question is the connection between low and high $Q^2$. For the hard 
piece there is no problem, as both expressions are identical up to 
NLO corrections. For the soft piece, if we 
start with a constant behaviour for $Q^2\sim 2\,-\,5\,\gev^2$, then as $Q^2$ 
grows an expression like (2) will start to develop. the details of 
this will depend on what one assumes for the gluon structure function. 
Because the variation both with $Q^2$ and $x$ of the soft piece is slower 
than that of the hard part, we think the best procedure 
is to assume constancy of the soft piece up to $Q^2=8.5\,\gev^2$, and 
the evolved form from there on; since a very good fit is obtained at the low 
momentum region already with the constant behaviour
 there is little point in adding frills, and a new 
constant (the soft component of the gluon structure 
function).

Next, we say a few words on the parameters, starting with the 
QCD parameter, $\Lambdav$. It is impossible to give a 
reliable determination of the value of this parameter from low $x$ data alone;
 if fitting it, the central 
values vary between $0.08\,\gev$ and $0.55\,\gev$. If we take what we consider 
the more reliable fits, those in Tables X{\it a, c},
 and allow $\Lambdav$ to vary, we find an optimum value of $0.31\,\gev$; 
in particular, from the fit with 
hard plus soft Pomerons, plus $P'$, the 
optimum is $\Lambdav=0.32\pm 0.05\;\mev$ for a \chidof\ of 
$\tfrac{261}{230-6}$, hardly 
improving the result reported in Table X$c$. This 
tells us little more than rough compatibility between the low $x$ and other 
determinations of $\Lambdav$.

With respect to other parameters, we can say that the fits give precise 
determinations of the {\it nonsinglet} parameters, $\rho,\,B_{NS}$; 
but the singlet parameters are much less precisely determined. For 
example, $\lambda$ varies from 0.38 (LO, hard singularity only) to 
0.32 (NLO, hard singularity only) to $0.47\pm 0.04$ (hard plus soft, 
the best value in our opinion). Likewise, $B_S$ varies by almost one order 
of magnitude. The reason for this may be traced to the dependence of 
the parameters on the 
theoretical formulas used to fit the data, in particular when going from LO to NLO because of  
the large size of the NLO corrections to $F_S$. We consider the parameters given
 in Tables X$a,\,c$ to be the more reliable ones in particular for extrapolations to larger $Q^2$.

\vskip.4cm
\centerline{\bf APPENDIX}
\vskip.5cm
In this Appendix we present the full collection of
 formulas necessary to evaluate electroproduction 
to NLO.

\noindent{\sl Leading order quantities.}

$\beta_0=11-\frac{2}{3}n_f,\,\beta_1=102-\frac{38}{3}n_f$.
$$d_{NS}(n)=-\frac{\gamma^{(0)}_{NS}(n)}{2\beta_0};\;
{\bf D}(n)=-\frac{\ybf{\gamma}^{(0)}(n)}{2\beta_0}.\eqno ({\rm A}.1)$$
Here, $\gamma^{(0)}_{NS}(n)=\gamma^{(0)}_{11}(n)$ and
$$\ybf{\gamma}^{(0)}=-\tfrac{32}{3}\left(\matrix{\dfrac{1}{2n(n+1)}+\frac{3}{4}-S_1(n)&
\kern-4em\frac{3}{8}n_f\;\dfrac{n^2+n+2}{n(n+1)(n+2)}\cr
\kern-4.4em\dfrac{n^2+n+2}{2n(n^2-1)}&
\kern-4.1em\dfrac{33-2n_f}{16}+\frac{9}{4}\,
\left[\dfrac{1}{n(n-1)}+\dfrac{1}{(n+1)(n+2)}-S_1(n)\right]\cr}\right)\eqno ({\rm A}.2)$$
We define, generally, the functions
$$S_l(n)=\sum^{\infty}_{k=1}\left[\frac{1}{k^l}-\frac{1}{(k+n)^l}\right],\;
 S^+_l(\tfrac{1}{2}n)=S_l(n/2);\; S^-_l(\tfrac{1}{2}n)=S_l\left(\dfrac{n-1}{2}\right),$$
and
$$\tilde{S}^{\pm}(n)=-\tfrac{5}{8}\zeta(3)\mp
\sum^{\infty}_{k=1}\frac{(-1)^k}{(k+n)^2}\;S_1(k+n).$$

The matrix that diagonalizes ${\bf D}(n)$ is ${\bf S}(n)$,
$${\bf S}(n)^{-1}{\bf D}(n){\bf S}(n)=\left(\matrix{d_+(n)&0\cr0&d_-(n)\cr}\right),$$
with the eigenvalues ordered so that $d_+>d_-$. ${\bf S}$ may be written as
$${\bf S}(n)=\left(\matrix{1&\dfrac{D_{12}(n)}{d_-(n)-d_+(n)}\cr
\dfrac{d_+(n)-D_{11}(n)}{D_{12}(n)}&
\dfrac{d_-(n)-D_{11}(n)^{\phantom{x}}}{d_-(n)-d_+(n)}\cr}\right).
\eqno ({\rm A}.3)$$

\noindent{\sl Nonsinglet NLO quantities.}

We will not give explicit formulas for the $C_{NS}^{(1)}(n), \,\gamma_{NS}^{(1)}(n)$, which 
 may be found, misprint free, in refs. 6, 7. We will only present a few 
 figures for relevant values of $n$ from which good interpolation formulas 
may be written. One has  
$$C_{NS}^{(1)}(n=1)=0;\;\gamma_{NS}^{(1)}(1)=\tfrac{8}{9}[13+8\zeta(3)-2\pi^2],$$
the last formula for crossing-even functions (like $F_{NS}$ in electroproduction). For 
crossing odd functions, like $xF_3$ 
in neutrino scattering, the corresponding NLO anomalous dimension coefficient 
 verifies 
$$\gamma_{NS,{\rm odd}}^{(1)}(1)=0.$$ 
For $n$ near 0.5,
\vskip.2cm
\setbox0=\vbox{\hsize14cm
\vskip.2cm

{\petit
$$\matrix{ &n=0.4&n=0.5&n=0.6\cr
C_{NS}^{(1)}(n)=&17.1;&9.6;&5.5\cr}$$
\vskip.1cm
\hrule
$$\matrix{& &n_f=3& & & &n_f=4& & & &n_f=5& \cr
n=&0.4&0.5&0.6;&&0.4&0.5&0.6;&&0.4&0.5&0.6\cr
\gamma_{NS}^{(1)}=&-273.7&-159.3&-98.2;&&-271.7&-155.9&-95.0;&&-269.6&-152.5&-91.7\cr}$$
}
\vskip.1cm
}
\centerline{\boxit{\box0}}

Note that $C_{NS}^{(1)}(n)$ is independent of $n_f$.

\noindent{\sl Singlet NLO quantities.}

The four quantities $C_{ij}(n)$ may be found in ref. 6. Here we give
 only the two that enter the 
calculation for $ep$ scattering. With $C_F=4/3, C_A=3, T_F=1/2$,
$$C^{(1)}_{11}(n)=C_F\left\{2S_1^2(n)+3S_1(n)-2S_2(n)-\dfrac{2S_1(n)}{n(n+1)}-
9+\dfrac{3}{n}+\dfrac{4}{n+1}+\dfrac{2}{n^2}\right\};\eqno ({\rm A}.4{\rm a})$$
$$C^{(1)}_{12}(n)=4n_fT_F\left\{-\dfrac{1}{n}+\dfrac{1}{n^2}+\dfrac{6}{(n+1)(n+2)}-
S_1(n)\,\dfrac{n^2+n+2}{n(n+1)(n+2)}\right\}.\eqno ({\rm A}.4{\rm b})$$
Finally, for the NLO anomalous dimension matrix $\ybf{\gamma}^{(1)}(n)$ we have
 the following expressions:
$$\eqalign{\gamma^{(1)}_{11}(n)=
 C_F^2\left\{16S_1(n)\,\dfrac{2n+1}{n^2(n+1)^2}+
16\left(2S_1(n)-\dfrac{1}{n(n+1)}\right)\big[S_2(n)-S_2^+(\tfrac{1}{2}n)\big]\right.\cr
 \left.+24S_2(n)+64\tilde{S}(n)-8S_3^+(\tfrac{1}{2}n)-3-8\,\dfrac{3n^3+n^2-1}{n^3(n+1)^3}
-16\,\dfrac{2n^2+2n+1}{n^3(n+1)^3}\right\}\cr
 +C_F C_A\left\{ \tfrac{536}{9}S_1(n)-8\left(2S_1(n)-
\dfrac{1}{n(n+1)}\right)\big[2S_2(n)-S_2^+(\tfrac{1}{2}n)\big]\right.\cr
-\tfrac{88}{3}S_2(n)-32\tilde{S}(n)+4S_3^+(\tfrac{1}{2}n)-\tfrac{17}{3}\cr
\left.-\tfrac{4}{9}\,\dfrac{151n^4+236n^3+88n^2+3n+18}{n^3(n+1)^3}+
8\,\dfrac{2n^2+2n+1}{n^3(n+1)^3}\right\}\cr
 +n_fT_FC_F\left\{-\tfrac{160}{9}S_1(n)+\tfrac{32}{3}S_2(n)+\tfrac{4}{3}\right.\cr
\left.\kern4em +\tfrac{16}{9}\,
\dfrac{11n^7+49n^6+5n^5-329n^4-514n^3-350n^2-240n-72}{(n-1)n^3(n+1)^3(n+2)^2}\right\};\cr}
\eqno ({\rm A}.5{\rm a})$$
\vskip.2cm
$$\eqalign{-\gamma^{(1)}_{12}(n)= 8n_fT_FC_A
\left\{\big[-2S_1^2(n)+2S_2(n)-2S_2^+(\tfrac{1}{2}n)\big]\,\dfrac{n^2+n+2}{n(n+1)(n+2)}\right.\cr
+8S_1(n)\,\dfrac{2n+3}{(n+1)^2(n+2)^2}+\dfrac{3n^4+15n^3+29n^2+50n+44}{n(n+1)^3(n+2)^3}\cr
 \left.+\dfrac{2n^9+12n^8+27n^7+38n^6+58n^5+149n^4+262n^3+252n^2+128n+32}
{(n-1)n^3(n+1)^3(n+2)^3}\right\}\cr
+8n_fT_FC_F\left\{\big[2S_1^2(n)-2S_2(n)+5\big]\dfrac{n^2+n+2}{n(n+1)(n+2)}-
\dfrac{4S_1(n)}{n^2}\right.\cr
\left.+\dfrac{11n^4+26n^3+15n^2+8n+4}{n^3(n+1)^3(n+2)}\right\}.\cr}\eqno ({\rm A}.5{\rm b})$$
This corrects a misprint in ref. 8 ( a figure $262n^3$ instead of $26n^3$ in the third line).
$$\eqalign{-\gamma^{(1)}_{21}(n)=
 4C_F^2 \left\{\big[-2S_1^2(n)+10S_1(n)-2S_2(n)\big]\dfrac{n^2+n+2}{(n^2-1)n}\right.\cr
\left.-\dfrac{4S_1(n)}{(n+1)^2}-
\dfrac{12n^6+30n^5+43n^4+28n^3-n^2-12n-4}{(n-1)n^3(n+1)^3}\right\}\cr
 +8C_FC_A\left\{\big[S_1^2(n)+S_2(n)-S_2^+(\tfrac{1}{2}n)\big]\,\dfrac{n^2+n+2}{n(n^2-1)}\right.\cr
 -S_1(n)\,\dfrac{17n^4+41n^2-22n-12}{3(n-1)^2n^2(n+1)}+
\dfrac{n^3+n^2+4n+2}{n^3(n+1)^3}\cr
\left.+\dfrac{109n^8+512n^7+879n^6+772n^5-104n^4-954n^3-278n^2+288n+72}
{9(n-1)^2n^3(n+1)^2(n+2)^2}\right\}\cr
+\tfrac{32}{3}n_fT_FC_F\left\{\big[S_1(n)-\tfrac{8}{3}\big]\,
\dfrac{n^2+n+2}{n(n^2-1)}+\dfrac{1}{(n+1)^2}\right\};\cr}
\eqno ({\rm A}.5{\rm c})$$
\vskip.2cm
$$\eqalign{\gamma^{(1)}_{22}(n)= n_f T_F C_A 
\left\{-\tfrac{160}{9}S_1(n)+\dfrac{32}{3}+
\tfrac{16}{9}\,\dfrac{38n^4+76n^3+94n^2+56n+12}{(n-1)n^2(n+1)^2(n+2)}\right\}\cr
+n_f T_F C_F \left\{ 8+
16\,\dfrac{2n^6+4n^5+n^4-10n^3-5n^2-4n-4}{(n-1)n^3(n+1)^3(n+2)}\right\}\cr
 +C_A^2\left\{-16S_1(n)S_2(n)+32\tilde{S}(n)-4S_3^+(\tfrac{1}{2}n)\right.\cr
 +32S_1(n)\left[\tfrac{67}{36}+\dfrac{1}{(n-1)^2}-\dfrac{1}{n^2}+
\dfrac{1}{(n+1)^2}-\dfrac{1}{(n+2)^2}\right]\cr
 +16S_2(n)\left[\dfrac{1}{n-1}-\dfrac{1}{n}+\dfrac{1}{n+1}-\dfrac{1}{n+2}\right]\cr
+16\big[S_2(n)-S_2^+(\tfrac{1}{2}n)\big]\,
\left[S_1(n)-\dfrac{1}{n(n-1)}-\dfrac{1}{(n+1)(n+2)}\right]\cr
 -\tfrac{64}{3}-\dfrac{32}{n-1}-\dfrac{148}{9n(n+1)}+\dfrac{32}{n+2}-\dfrac{32}{(n-1)^2}\cr
\left.-\dfrac{8}{3n^2}+\dfrac{88}{3(n+1)^2}-\dfrac{256}{3(n+2)^2}
-\dfrac{32}{n^3}-\dfrac{32}{(n+1)^3}-\dfrac{64}{(n+2)^3}\right\}\cr
}\eqno ({\rm A}.5{\rm d})$$
\vskip.2cm
This last corrects two misprints of ref. 8, a factor $n_fT_F\equiv T_R$ instead of $T_A$ 
in the first line, and a sign, $+1/(n+1)^2$ instead of $-1/(n+1)^2$ in the fourth line.

We finish by giving two values of the $C^{(1)}_{ij},\,\gamma^{(1)}_{ij}$  
and tables with a few listings, sufficient for the
 calculations we are interested in. As for the first, we have
$$C^{(1)}_{11}(2)=\tfrac{4}{9},\;C^{(1)}_{12}(2)=-\tfrac{1}{2}n_f,$$
$$\ybf{\gamma}^{(1)}(2)=\tfrac{64}{243}
\left(\matrix{367-39n_f&-\tfrac{1833}{32}n_f\cr
-(367-39n_f)&\phantom{\dfrac{1}{1}}\tfrac{1833}{32}n_f\cr}\right).$$

We give the tables for the quantities relevant for the exponential expression, 
 $q_S$ and $c_S$,
$$\eqalign{F_S(x,Q^2)\simeqsub_{x\rightarrow 0
}B_S\left\{1+\frac{c_S(1+\lambda)\alpha_s(Q^2)}{4\pi}\right\}\cr
\times\left[\alpha_s(Q^2)\right]^{-d_+(1+\lambda)}
\ee^{q_S(1+\lambda)\alpha_s(Q^2)/4\pi}x^{-\lambda},}$$
with
$$c_S=C^{(1)}_{11}+\frac{(d_+-D_{11})C^{(1)}_{12}}{D_{12}},$$
$$q_S=\frac{\beta_1d_+}{\beta_0}+\frac{\bar{\gamma}_{11}}{2\beta_0}+
\frac{\bar{\gamma}_{21}}{2\beta_0(d_--d_++1)}\;\frac{D_{12}}{d_--d_+};$$
for the expanded expression just
note that $w_S=q_S+c_S$. Then,
\vskip.2cm
\setbox0=\vbox{\hsize=14cm
\vskip.15cm
{\petit
$$\matrix{ &1+\lambda=&1.20 &1.275&1.3 &1.325&1.35&1.375&1.42&1.47&1.50&1.94\cr
& & & & & & & & & \cr
n_f=3&q_S&142.2&99.9&91.0&83.7&77.8&72.9&66.3&61.8&60.5&-5.7\cr
n_f=4&q_S&147.9&103.1&93.5&85.7&79.2&73.8&66.1&60.2&57.8&-13.8\cr
n_f=5&q_S&153.6&106.2&96.1&87.6&80.6&74.6&66.1&59.0&55.8&-82.0\cr
& & & & & & & & & & \cr
n_f=3&c_S&8.28&1.74&0.44&-0.59&-1.41&-2.07&-2.93&-3.53&-3.76&-2.61\cr
n_f=4&c_S&8.26&1.73&0.43&-0.59&-1.41&-2.06&-2.91&-3.50&-3.72&-2.59\cr
n_f=5&c_S&8.23&1.72&0.43&-0.59&-1.42&-2.05&-2.89&-3.47&-3.69&-2.57\cr}$$
}
\vskip.1cm
}
\centerline{\boxit{\box0}}

For the gluon structure function,
$$\eqalign{F_G(x,Q^2)\simeqsub_{x\rightarrow 0}
 B_S\frac{d_+(1+\lambda)-D_{11}(1+\lambda)}{D_{12}(1+\lambda)}\cr
\times\left\{1+\frac{c_G(1+\lambda)\alpha_s(Q^2)}{4\pi}\right\}
\left[\alpha_s(Q^2)\right]^{-d_+(1+\lambda)}
\ee^{q_G(1+\lambda)\alpha_s(Q^2)/4\pi}x^{-\lambda},}$$
with
$$c_G\equiv c_S,$$
$$q_G=\frac{\beta_1d_+}{\beta_0}+\frac{\bar{\gamma}_{11}}{2\beta_0}+
\frac{\bar{\gamma}_{21}}{2\beta_0(d_- -d_++1)}\;\frac{D_{12}}{d_--d_+}\;
\frac{d_--D_{11}}{d_+-D_{11}}.$$
One has,
\vskip.2cm
\setbox0=\vbox{\hsize=14cm
\vskip.2cm
{\petit
$$\matrix{ &1+\lambda=&1.20 &1.275&1.3 &1.325&1.35&1.375&1.42&1.47&1.50&1.94\cr
& & & & & & & & & \cr
n_f=3&q_G&37.59&20.43&16.77&13.74&11.17&8.96&5.65&2.61&0.96& \cr
n_f=4&q_G&42.49&23.04&18.87&15.39&12.44&9.89&6.06&2.56&0.69& \cr
n_f=5&q_G&47.36&25.60&20.91&16.99&13.65&10.77&6.44&2.51&0.43&\cr}$$
}
\vskip.1cm
}
\centerline{\boxit{\box0}}

\noindent{\sl Coefficients and integrals for the longitudinal structure function.}

We give the coefficient functions for $ep$ 
scattering, with unified notation. With the definitions of Eqs. (2.12), and with 
the polylogarithm functions\fonote{$L_{1,2}$ denoted $S_{1,2}$ in refs. 12, 13.}
$${\rm Li}_{\nu}(x)=\sum_{n=1}^{\infty}\frac{x^n}{n^{\nu}};\;
L_{1,2}(x)=\tfrac{1}{2}\int^1_0\dd t\,\frac{\log^2(1-tx)}{t},$$
we have,
$$\eqalign{c_{NS}^{(1)L}(x)=4C_F(C_A-2C_F)
\bigg\{4\frac{6-3x+47x^2-9x^3}{15x}\log x\cr
-4x{\rm Li}_2(-x)\big[\log x-2\log (1+x)\big]\cr
-2x\log^2x\log(1-x^2)+4x\log x\log^2(1+x)-4x\log x\,{\rm Li}_2(x)\cr
+2x(1-\tfrac{3}{5}x^2)\log^2 x-\frac{144+294x-1729x^2+216x^3}{90x}\cr
-4\frac{2+10x^2+5x^3-3x^5}{5x^2}\big[{\rm Li}_2(-x)+\log x\log(1+x)\big]\cr
+4x\zeta(2)\big[\log(1-x^2)-1+\tfrac{3}{5}x^2\big]-8x\zeta(3)\cr
+8xL_{1,2}(-x)+4x[{\rm Li}_3(x)+{\rm Li}_3(-x)]-\tfrac{23}{3}x\log(1-x)\bigg\}\cr
+8C_F^2\bigg\{x{\rm Li}_2(x)+x\log^2\frac{x}{1-x}-3x\zeta(2)\cr
-\left(1-\tfrac{22}{3}x\right)\log x+
\left(1-\tfrac{25}{6}x\right)\log(1-x)-\frac{78-355x}{36}\bigg\}\cr
-\tfrac{16}{3}n_fT_FC_F\bigg\{x\log\frac{x^2}{1-x}-1+\tfrac{25}{6}x\bigg\};
}\eqno ({\rm A.6a})$$
\vskip.1cm
$$\eqalign{c_{PS}^{(1)L}(x)=\tfrac{32}{9}n_fT_FC_F\bigg\{3\frac{1-2x-2x^2}{x}(1-x)\log (1-x)
-9x(1-x)\cr
-\frac{(1-x)^3}{x}+9x\big[{\rm Li}_2(x)+\log^2x-\zeta(2)\big]+9(1-x-2x^2)\log x\bigg\};
}\eqno ({\rm A.6b})$$
\vskip.1cm
$$\eqalign{c_{G}^{(1)L}(x)=2n_fT_FC_F
\bigg\{(8+24x-32x^2)\log(1-x)\cr
+16x\big[{\rm Li}_2(1-x)+\log x\log(1-x)\big] \cr
+\left(-\tfrac{32}{3}x+\tfrac{64}{5}x^3+\tfrac{32}{15}x^{-2}\right)
\big[{\rm Li}_2(-x)+\log x\log(1+x)\big]\cr
+\frac{\log x}{15}\left[-104-624x+288x^2-32x^{-1}\right]
-32x\left(\tfrac{1}{3}+\tfrac{1}{5}x^2\right)\log^2x\cr
+\left(-\tfrac{32}{3}x+\tfrac{64}{5}x^3\right)\zeta(2)
-\tfrac{128}{15}-\tfrac{304}{5}x+\tfrac{336}{5}x^2+\tfrac{32}{15}x^{-1}\bigg\}\cr
+2n_fT_FC_A\bigg\{-64x{\rm Li}_2(1-x)+32x(1+x)\big[{\rm Li}_2(-x)+\log x\log(1+x)\big]\cr
+16x(1-x)\log^2(1-x)+x(-96+32x)\log x\log(1-x)+48x\log^2x\cr
+32x^2\zeta(2)+[16+128x-208x^2]\log x
+\tfrac{16}{3}+\tfrac{272}{3}x-\tfrac{848}{9}x^2-\tfrac{16}{9}x^{-1}\cr
+\left[-16-144x+\tfrac{464}{3}x^2+\tfrac{16}{3}x^{-1}\right]\log(1-x)\bigg\}.
}\eqno ({\rm A.6c})$$

Integrals: for $n_f=4$,
\vskip.2cm
\setbox0=\vbox{\hsize=14cm
\vskip.2cm

{\petit
$$\matrix{ \lambda=&0&0.25&0.30&0.325&0.35&0.375&0.42&0.47&0.50\cr
 & & & & & & &\cr
I_{S1}=&15.4&15.3&15.3&15.2&15.1&15.1&15.0&14.9&14.9\cr
I_{S2}=&30.4&29.3&29.3&29.2&29.2&29.2&29.2&29.2&29.1\cr
I_{S3}=&-7.11&-7.28&-7.28&-7.28&-7.27&-7.26&-7.25&-7.23&-7.21\cr
 & & & & & & & &  &\cr
I_{PS}=& &-22.8&-20.1&-18.9&-17.8&-16.8&-15.3&-13.8&-13.0\cr
 & & & & & & & & &\cr
I_{G1}=&-10.7&-11.4&-11.4&-11.3&-11.3&-11.2&-11.2&-11.0&-11.0\cr
I_{G2}=& &-2.42&0.30&1.46&2.52&3.48&4.99&6.41&7.14\cr}$$
}
\vskip.1cm
}
\centerline{\boxit{\box0}}
\vskip.3cm
(the values of the integrals not given when $x$-dependent).

\vfill\eject
\centerline{ \bf ACKNOWLEDGEMENTS}
\vskip .5cm
We acknowledge interesting discussions with J. Terr\'on.
 One of us (FJY) is grateful to J. S\'anchez-Guill\'en and
 G. Parente for information concerning the NLO corrections to the longitudinal structure 
functions, to A. B. Kaidalov for the relevance of multi-Regge 
analysis, and to M. McDermott for communicating the 
results of calculations which prompted 
the detailed study of the NLO predictions of the soft Pomeron model. 

This work was partially supported by CICYT, Spain.
}
\vskip.4cm 
\centerline{\bf REFERENCES}
\vskip.5cm
{\petit
\noindent 1. F. J. Yndur\'ain,Preprint FTUAM 96-12 (revised),
 to be published in Proc. QCD 96, Nucl. Phys. Suppl.\hb
2. F. Barreiro, C. L\'opez and F. J. Yndur\'ain, \jzp{72}{1996}{561}. \hb
3.- M. Derrick et al, Z. Phys.C, {\bf 65} (1995) 397; Phys. Lett. {\bf B345} 
(1995) 576.\hb
4. H1 Collaboration, preprint DESY 96-039, 1996; 
Zeus Collaboration, preprint DESY 96-076, 1996.\hb
5. C. L\'opez and F. J. Yndur\'ain, Nucl. Phys., {\bf B171} (1980) 231 (LO).\hb 
6. C. L\'opez and F. J. Yndur\'ain, Nucl. Phys., {\bf B183} (1981) 157 (NLO).\hb
7. F. J. Yndur\'ain, {\sl Quantum Chromodynamics}, Springer 1983; second 
edition as {\sl The Theory of Quark and Gluon Interactions}, Springer. 1992\hb
8. A. Gonz\'alez-Arroyo, C. L\'opez and F. J. Yndur\'ain, Phys. Lett., {\bf 98B} (1981) 215.\hb
9. A. De R\'ujula et al., Phys. Rev., {\bf D10} (1974) 1649. \hb
10. F. Martin, Phys. Rev., {\bf D19} (1979) 1382.\hb
11. R. D. Ball and S. Forte, Phys. Lett., {\bf B358} (1995) 365 and CERN TH/95-323 (1995), 
to be published in Acta Phys. Polonica.\hb
12. J. S\'anchez-Guill\'en et al., Nucl. Phys., {\bf B353} (1991) 337.\hb
13. E. B. Zijstra and W. L. van Neerven , Phys. Lett., {\bf B272} (1991) 476.\hb  
14. S. A. Larin and J. Vermaseren, Z. Phys. C, {\bf 57} (1993) 93.\hb 
15. E. Oltman et al., Z. Phys. C, {\bf 53} (1992) 51; 
 P. Berge et al., Z. Phys. C, {\bf 49} (1991) 187; 
 P. Z. Quintas et al., Phys. Rev. Lett., {\bf 71} (1993) 1307.\hb
16. Particle Data Group, any issue.\hb
17. M. Klein (H1 collaboration), 4th Int. Conf. on Deep inelastic Scattering, Rome, 1996.\hb
18. H. L. Anderson et al., Phys. Rev., {\bf D20} (1979) 2645; A. Bodek et al. 
{\it ibid}, 1471.\hb
19. E. A. Kuraev, L. N. Lipatov and V. S. Fadin, Sov. Phys. JETP, {\bf 44} (1976) 443; 
Ya. Ya. Balitskii and L. N. Lipatov, Sov. J. Nucl. Phys., {\bf 28} (1978) 822.\hb 
20. M. Ciafalloni, Nucl. Phys. {\bf B296} (1988) 49; 
S. Catani, F. Fiorini and G. Marchesini, Phys. Lett., {\bf B234} (1990) 339 
and Nucl. Phys. {\bf B336} (1990) 18; 
S. Catani and F. Hautmann, Nucl. Phys., {\bf B427} (1994) 475;
 J. Bl\"umlein and A. Vogt, Acta Phys. Polonica, {\bf B27} (1996), 1309.\hb
21. A. De Roeck, M. Klein and Th. Naumann, DESY 96-063 (1996).\hb
22. C. L\'opez and F. J. Yndur\'ain, Phys. Rev. Lett., {\bf 44} (1980) 1118.\hb
23. E. Witten, Nucl. Phys., {\bf B119} (1977) 189.\hb
24. A. B. Kaidalov, Survey in High Energy Physics, Vol. 9 (1996) 143.\hb
25. V. D. Barger and D. B. Cline, {\sl Phenomenological Theories 
of High Energy Scattering}, Benjamin, 1969.\hb
26. H. Klein, DESY 96-218 (1996).\hb
27. M. Derrick et al., \jpl{350}{1995}{134}, and work quoted there.\hb
28. Yu. A. Simonov, Yadernaya Fizika, {\bf 58} (1995) 113, and work quoted there.\hb
29. B. Badelek and J. Kwiecinski, Phys. Lett. {\bf B295} (1992) 263; Rev. Mod. Phys.,
 {\bf 68} (1996) 445.\hb}

\bye